\documentclass[epj]{svjour}
\usepackage{graphics}
\begin{document}
\title{Eikonal representation in the momentum-transfer 
space}
\author{P.A.S. Carvalho\inst{1}, A.F. Martini\inst{2}, M.J. Menon\inst{2}}

\mail{menon@ifi.unicamp.br}          
\institute{Centro Universit\'ario de Patos de Minas, 38702-054 Patos de 
Minas, MG, Brazil
\and 
Instituto de F\'{\i}sica Gleb Wataghin
Universidade Estadual de Campinas, 13083-970 Campinas, SP Brazil}
\date{Received: date / Revised version: date}
%
\abstract{
By means of empirical fits to the differential cross section data on $pp$ 
and
$\bar{p}p$ elastic scattering, above 10 GeV (center-of-mass energy), we 
determine 
the eikonal in
the momentum - transfer space ($q^2$- space). We make use of a numerical 
method and
a novel semi-analytical method, through which the uncertainties from 
the fit parameters 
can be propagated up to the eikonal in the $q^2$- space. A systematic 
study of the
effect of
the experimental information at large values of the momentum transfer is
developed and discussed in detail. 
We present statistical evidence that the imaginary part of the eikonal 
changes 
sign in the 
$q^2$- space and that the position of the zero decreases as the energy 
increases;
after the position of the zero, the eikonal presents a minimum and then 
goes to zero through negative values.
We discuss the applicability of our results in the phenomenological context, 
outlining some connections with nonperturbative QCD. 
A short review and a critical discussion on the
main results concerning ``model-independent" analyses are also presented.
} 
\maketitle
\section{Introduction}
\label{intro}

High-energy particle scattering is the main experimental tool in
the investigation of the inner structure of matter. For particle-particle
and antiparticle-particle scattering, the highest energies reached in
accelerators concern proton-proton ($pp$) and antiproton-proton 
($\bar{p}p$) collisions and these hadronic processes are expected 
to be described 
by the Quantum Chromodynamics (QCD), the gauge invariant quantum 
field theory 
of the strong interactions.
Recently, Dokshitzer stated that ``QCD nowadays has a split personality.
It embodies hard and soft physics, both being hard subjects and the
softer the harder" \cite{dok}. In fact, despite of all the success of
QCD in the treatment of hard and semi-hard scattering processes
(large and medium momentum transfer, respectively), the increase of the
coupling constant when going to  the soft sector (small momentum transfer)
does not allow the use of the most important tool in theoretical physics,
namely the perturbative calculation.
On the other hand, the nonperturbative QCD approach, characterized by the 
non-trivial investigation of the vacuum structure and intricate simulation 
techniques in the study of bound states, did not yet provide any result for
the scattering states, based exclusively on its own foundations.
Soft scattering embodies diffraction dissociation (single and double) and
elastic scattering \cite{pred} and the point is that, presently, we do 
not know how 
to calculate even the elastic scattering amplitude (the simplest soft process)
in a purely nonperturbative context. Certainly, there are QCD-based or
QCD-inspired models, but, anyway, they have only bases or inspiration in
QCD \cite{martin}.

However, from the experimental point of view, the expectations 
in the field are
very high due to the new generation of colliders. From experiments that are 
being conducted at the BNL Relativistic Heavy Ion Collider (RHIC) 
it is expected data on $pp$ soft scattering 
at center-of mass energies $\sqrt s$: 50 - 500 GeV, and in the near future 
the CERN 
Large Hadron Collider (LHC) will 
provide data on soft $pp$ scattering at 14 TeV. 

At this stage, beyond the efforts directed to pure theoretical developments
(nonperturbative QCD) and phenomenology, 
\textit{``empirical" analyses} play an important role in the search for
model-independent results that can contribute with the establishment
of novel and useful theoretical calculational schemes.
In order to achieve that, any approach must be based on the
General Principles and Theorems of the underlying field theory,
namely Unitarity, Analyticity, Crossing and their well founded consequences.
Among the wide variety of formalisms \cite{matthiae}, the
\textit{eikonal representation} is distinguished by its intrinsic connection
with Unitarity \cite{valin} and its efficiency as a useful framework for
phenomenological analyses. Eikonal models can be distinguished by the different
forms of the eikonal in the \textit{momentum-transfer space}
($q^2$ - space); for
our purposes, we list some important and well known results in Refs.
\cite{glauber,henzivalin,chouyang,gv,bsw,valinetal,bghp}.
In this context, analyses based on empirical fits to 
the physical quantities involved and aimed to extract the characteristics of 
the eikonal as function of the momentum transfer, energy and reaction play an
essential role in the attainment of consistent results in a 
\textit{model-independent} way.

 In a previous work \cite{cm} we have investigated $pp$ scattering at the CERN 
Intersecting Storage Ring (ISR) energy region, namely $\sqrt s =$
23.5, 30.7, 44.7, 52.8 and 62.5 GeV. By means of fits to the differential cross
section data and through a novel ``semi-analytical" method,
which allowed the error propagation from the fitted parameters,
we have determined the imaginary part of the eikonal in the momentum transfer 
space and have found
statistical evidence for the existence of eikonal zeros
(change of sign) in the region of momentum transfer $q^2 =$ 7 $\pm$ 2 GeV$^2$.
These results have been used and discussed 
in different contexts \cite{kawa1,kawa2,fp,mmt}.

In the present work, we have extended our previous analysis in several ways: 
(1) we include the 
$pp$ data at lower energies, namely $\sqrt s =$ 13.8 GeV and 19.5 GeV and
also $\bar{p}p$ data in the region
14 GeV $\leq \sqrt s \leq$ 1.8 TeV; (2) we make use of a numerical calculation
in order to check the results obtained with the ``semi-analytical"
 method; (3) we present in detail the statistical
regions of confidence and all the numerical information; (4) 
we discuss the applicability of the results in the phenomenological 
context and also some possible connections with nonperturbative QCD
results. In addition, we present a critical discussion on the
different analyses aimed to extract the eikonal from fits to the
differential cross section data.
Our main 
novel conclusions
are that the position of the eikonal zero  
decreases as the energy increases and, after the zero, the eikonal has 
a minimum and then goes to zero through negative values. 

The manuscript is organized as follows. In Sect. \ref{sec:2} we 
treat the Eikonal representation,
recalling the formulas connecting the physical quantities 
that characterizes the elastic scattering with
the eikonal in the $q^2$- space and stressing the importance of the
``empirical" information. In Sect. \ref{sec:3}, after introducting the
experimental data to be analyzed, we
shortly review some typical
results and open problems associated with model-independent analyses
and, based on these considerations, we present our
strategies, mainly related to
the definition of two ensembles of experimental information.
In Sect. \ref{sec:4} we display the results of the fits to the differential 
cross section data, discussing in detail our choice for the
parametrization, the fit procedure and the effect of the
uncertainties.
In Sect. \ref{sec:5} we treat the determination of the eikonal in
the momentum transfer space by means of both the
numerical and semi-analytical methods. 
In Sect. \ref{sec:6} we discuss 
the applicability of our results in a phenomenological 
context and possible connections with nonpertubative QCD 
results. 
The conclusions and some
final remarks are the contents of Sect. \ref{sec:7}.

\section{Eikonal Representation}
\label{sec:2}

In this Section we first recall the essential formulas connecting
the eikonal in the momentum-transfer space and the physical
quantities to be investigated. Next, we discuss the importance of the
model-independent analyses, in particular some questions related with
eikonal zero (change of sign in the $q^2$ - space).

\subsection{ General formalism}

In the Eikonal Representation, the elastic scattering
amplitude, as function of the center-of-mass energy, $\sqrt s$, and
the four-momentum transfer squared, $q^2 \equiv - t$, is expressed by

\begin{eqnarray}
F(s, q) = i \int_{0}^{\infty} bdb J_{0}(qb)
\{ 1 - e^{i\tilde\chi(s,b)} \},
\end{eqnarray}
where $b$ is the impact parameter, $J_0$ is the zero order Bessel function 
(azimuthal symmetry assumed) and 
$\tilde\chi(s,b)$ is the eikonal function in the
impact parameter space. Although this formula may be deduced,
for example, from the partial wave solution of the Schroe\-din\-ger 
equation in the
high energy limit \cite{glauber}, it should be stressed its character 
of well founded
mathematical \textit{representation}, independent of any approximation. 
In fact, as
demonstrated by M.  Islam, the complex scattering amplitude $F(s,q)$
has a well defined representation in the impact parameter space,
valid for all energies and scattering angles \cite{islam}. This 
representation is
usually named Profile function and denoted $\Gamma(s,b)$. Since the
exponential function is an entire function of its argument,
we may always express the Profile function in terms of an eikonal
function:

\begin{eqnarray}
\Gamma (s,b) =  1 - e^{i\tilde\chi(s,b)}. 
\end{eqnarray}

Unitarity allows to connect elastic and inelastic processes, which, 
in the
impact parameter space, may be expressed by the formula 
\cite{valin,henzivalin}

\begin{eqnarray}
2 \mathrm{Re}\ \Gamma(s,b) = | \Gamma(s,b) |^2 + G_{inel}(s,b),  
\end{eqnarray}
where $G_{inel}(s,b)$ is the Inelastic Overlap Function, namely the 
probability for
an inelastic event to take place at $s$ and $b$. This probabilistic 
interpretation 
demands that 

\begin{eqnarray}
G_{inel}(s,b) \leq 1, \nonumber
\end{eqnarray}
and, from Eqs. (2) and (3), it is automatically 
verified for

\begin{eqnarray}
\textrm{Im}\ \tilde\chi(s,b) \geq 0.
\end{eqnarray}
This is the main result that characterizes the Eikonal representation,
 that is,
the automatic agreement with Unitarity, a principle that, certainly, can
never be violated.

The Eikonal approach has a long history of successful results that allowed
important developments. For example, from the earlies Chou-Yang 
\cite{chouyang} and
Glauber \cite{glauber,gv} models, passing through the interesting 
geometrical
formalism by Bourrely, Soffer and Wu \cite{bsw} and, more recently, the 
suggestive connections with quarks and gluons in the QCD-based models
\cite{valinetal,bghp}, the Eikonal picture seems to represent a nearly 
natural framework and suitable phenomenological laboratory.

All these eikonal models (and many others) may be 
distinguished or
classified according to different forms for the \textit{eikonal
function in the momentum transfer space}, which we shall represent
by $\chi(s,q)$, the Fourier-Bessel transform of $\tilde\chi(s,b)$:

\begin{eqnarray}
\chi(s,q) =  \int_{0}^{\infty} bdb J_{0}(qb) \tilde\chi(s,b).
\end{eqnarray}
In general the inputs for $\chi(s,q)$ are based on analogies 
with other areas (optics, geometry,...) and/or related with some
microscopic concepts (elementary processes), looking always for
the expected connection with quarks and gluons, in a well established
QCD bases.

The crucial test for any input comes from the experimental
data on the physical quantities that characterize the elastic
scattering. This demands going from Eq. (5) to (1) and then
to the differential cross section,

\begin{equation}
\frac{d\sigma}{dq^2} = \pi|F(s, q^2)|^2,
\end{equation}
the total cross section (Optical Theorem),

\begin{equation}
\sigma_{tot} = 4\pi \textrm{Im}\ F(s, q^2=0),
\end{equation}
the $\rho$ parameter

\begin{equation}
\rho = \frac{\textrm{Re}\ F(s, q^2=0)}{\textrm{Im}\ F(s, q^2=0)},
\end{equation}
and other quantities \cite{pred,matthiae}.

Comparisons with the corresponding data allow to check the ideas and, in a
feedback process, to reconsider concepts and suitable phenomenological
inputs, leading to new tests. However, to reach global and efficient
descriptions with an economical number of free parameters is a very
difficult task, even in the phenomenological context. One reason is because 
the physical quantities depend on both real and imaginary parts of
the scattering amplitude and those, in turn, depend on the energy, momentum
and reaction. Another reason concerns the fact that the model free
parameters are, in general, correlated in an intricate way, turning it
difficult to identify the explicit role or function of each parameter in the
description of the experimental data. Moreover, in most cases, one is
involved with numerical calculation, which introduces bias, loss
 of information,
and does not allow standard error propagation.

Certainly, the above mentioned problems can be avoided or, at least,
 minimized, if we
have some kind of ``empirical" or ``model-independent" information on the
eikonal directly in the $q^2$ - space. Once having well
established statistical bases, this information may provide suitable 
criteria
for input selections at the early stages of the model construction or
development. This ``inverse problem" concerns the determination of the
eikonal from fits to the experimental data and that is the point we are
interested in.

\subsection{Eikonal zeros in the momentum-transfer space}

One aspect that exemplifies the importance of this ``inverse problem"
is the possibility to extract model-indepen\-dent information on the
existence of eikonal zeros in the $q^2$ - space. The point is that the
diffraction minimum in the differential cross section (the dip) is 
generally interpreted as being associated with a zero in the
imaginary part of the scattering amplitude ($q^2$ - space) and, therefore,
from Eqs. (1) and (5-6), it seems natural to ask if this zero in the
amplitude is connected with, or is a consequence of, a zero in
the imaginary part of the eikonal in the $q^2$ - space. Moreover,
if that is the case, what is the phenomenological interpretation of the
zero? Since that is one of the main points we are interested to
discuss in this work, and also for future reference, let us shortly review,
in chronological order, some previous results on eikonal zeros, as well as
some phenomenological implications (see also \cite{kawa2}). 

By the end of the sixties, the ``coherent droplet model" introduced the
 idea that
the eikonal in the $q^2$-space could be expressed by the product of the
hadronic form factors, which, in turn, could be assumed similar to the
electromagnetic form factors \cite{chouyang}. Certainly, the similarity was 
thought in 
general geometrical terms, such as extents and smoothness. In this context,
the imaginary part of the factorized eikonal for $pp$ scattering reads

\begin{eqnarray}
\mathrm{Im}\ \chi(s,q) = C(s) G_{p}(q) G_{p}(q),
\end{eqnarray}
where $C(s)$ is an absorption coefficient and the proton form
factor was represented by the dipole parametrization for the
electromagnetic form factor:

 \begin{eqnarray}
 G_{p}(q) = \frac{1}{(1 + q^2/\mu^2)^2}, 
\qquad
\mu^2 = 0.71\ \textrm{GeV}^2.
\end{eqnarray}
Therefore, in this model $\mathrm{Im}\ \chi(s,q)$ is positive
(no change of sign) and the zero in the scattering
amplitude has been interpreted as an interference effect involving
both protons and not associated with the individual matter distribution
\cite{durandlipes}. 

However, in 1975, Victor Franco presented a 
detailed fit to $pp$ differential cross section data at $\sqrt s$ = 53 GeV
and $0 < q^2 \leq$ 5.3 GeV$^2$ \cite{franco}, showing that $\mathrm{Im}\ \chi$ 
may become negative for
$q^2 \geq $ 6.5 GeV$^2$ and concluding that the droplet model should be
modified. By that time there was also indication of zeros
in the pion form factor \cite{dubnicka} and in the proton form factor
\cite{bsw80}. Moreover, in 1977, by means of a multipole parametrization, 
Maehara, Yanagida and Yonezawa also obtained indication of a zero in  
the imaginary part of the eikonal
at $q^2 $= 6.0 GeV$^2$, from $pp$ scattering at $\sqrt s $ = 53 GeV \cite{myy}.

From a phenomenological or model point of view, the problem to introduce
an eikonal zero has been nicely resolved by Bourrely, Soffer and Wu (BSW)
in 1979,
by means of the following parametrization for the form factors in the
 $q^2$ - space
\cite{bsw,bsw02}:

\begin{eqnarray}
& & [G(q)]^2\  f_{\mathrm{BSW}}(q) =    \nonumber \\
& &  \left[ \frac{1}{(1 + q^2/m_1^2)} \frac{1}{(1 + q^2/m_2^2)} \right]^2 \
\left[\frac{1 - q^2/a^2}{1 + q^2/a^2}\right],
\end{eqnarray}
where $m_1$, $m_2$ and $a$ are free parameters. As referred in \cite{bsw02},
the product of two simple poles represents the ``nuclear form factor" and
the function $f_{\mathrm{BSW}}(q)$ ``reflects the approximate proportionality 
between the charge density and the hadronic matter distribution inside a proton".
With the implemented ``impact picture", BSW obtained 
good descriptions of the experimental
data with the zero fixed at $a^2 \approx$ 3.81 GeV$^2$ \cite{bsw} and,
more recently, at
$a^2 \approx$ 3.45 GeV$^2$ \cite{bsw02}.

In the eighties, by means of fits to $pp$ data, Sanielevici and Valin obtained 
indication of a zero in imaginary part of the eikonal at $q^2 \sim$ 5.0 GeV$^2$ 
\cite{sanivalin}
and latter, making use of the Amaldi and Schubert parametrization \cite{as},
Furget, Buenerd and Valin showed that the position of the zero decreases from
$q^2 \sim$ 8.6 GeV$^2$ to $\sim$ 5 GeV$^2$ as the energy increases from
$\sqrt s$ = 23.5 GeV to 62.5 GeV \cite{fbv}. 

As commented in our introduction, in 1997, Carvalho and Menon presented
statistical evidence for eikonal zeros from analyses of $pp$ scattering in the
limited energy interval 23.5 $\le \sqrt s \le$ 62.5 GeV$^2$. The analysis did
not allow to infer an energy dependence, but an estimated position of
the zero at $q^2$ = 7 $\pm$ 2 GeV$^2$ \cite{cm}.

Recently, Kawasaki, Maehara and Yonezawa investigated the connections
 between the zeros
of the scattering amplitude and the zeros of the eikonal in the
$q^2$ - space \cite{kawa2}. In a detailed study, the authors considered
several parametrizations for the scattering amplitude and introduced
interesting correlations between the eikonal zeros and a zero trajectory in the
$\sigma_{tot}$ versus $q^2$ plots. One of the conclusions of the work is the
indication of an eikonal zero at $q^2 \approx$ 7 GeV$^2$ (in
agreement with their previous results \cite{kawa1,myy}) and possibly no other
zeros in the region below $q^2 \approx$ 20 GeV$^2$.

From the above short review we conclude that there are indications of an 
eikonal zero 
at $q^2 \approx $ 7 GeV$^2$, which, in the phenomenological context, can 
be associated
with a zero in the hadronic form factor of the proton. However, despite 
the 
importance of this result, the analyses we referred to present one or 
another
kind of limitation, as we shall discuss in detail in  Sec. \ref{sec:3}.
 As a consequence,
the exact position of the zero is not yet clear and, most importantly, 
the possible
dependence of the position of the zero with the energy remains an open 
problem. 
To answer these and other questions (Sec. 3.2), by means of an improved
model-independent extraction of the eikonal, is the aim of this work.

\section{Experimental data, Strategies and Ensembles}
\label{sec:3}
In this Section we first refer to the experimental data to be
analyzed and recall the main problems and limitations related with
model-independent extraction of the eikonal from fits to the
differential cross section data. Based on this discussion, we
introduce novel procedures and strategies through which some of these
problems can be resolved or minimized, as demonstrated in the
Sections that follow.

\subsection{Experimental data}

As mentioned before, for particles and antiparticles the
$pp$ and $\bar{p}p$ scattering correspond to the highest energy interval 
with available data. In the high energy region, $\sqrt s > $10 GeV,
data on $\sigma_{tot}$, $\rho$ and differential cross section are available at 
$\sqrt s =$ 13.8, 19.5, 23.5, 30.7, 44.7, 52.8 and 62.5 GeV for $pp$
scattering \cite{as,schubert,pp,pppbarp} and at $\sqrt s =$ 13.8, 19.4,
 31, 53, 62, 546  and 
1800 GeV for $\bar{p}p$ scattering \cite{pppbarp,pbarp}.
Moreover,
differential cross section data from $pp$ scattering also exists at
$\sqrt s =$ 27.5 GeV 
and 5.5 $\leq q^2 \leq$ 14.2 GeV$^2$ \cite{schubert} and 
as discussed in the next Subsection, that
set will play a fundamental role in our analysis.

We have selected the differential cross section data above the region
of Coulomb-nuclear interference, namely $q^2 >$ 0.01 GeV$^2$ and the
data at $q^2 =$ 0 (optical point) is determined from the corresponding values
of $\sigma_{tot}$ and $\rho$:

\begin{eqnarray}
\left.\frac{d\sigma}{dt} \right|_{t=0} = 
\frac{\sigma_{tot}^2 (1 + \rho^2)}{16\pi}.
\end{eqnarray}

In terms of the momentum transfer, beyond the optical point, the $pp$ 
data cover the extended region 0.01 GeV$^2 < q^2 \leq $ 9.8 GeV$^2$
(except for the data at 27.5 GeV);
in contrast $\bar{p}p$ data are available only at 
0.02 GeV$^2 < q^2 \leq $ 4.45 GeV$^2$.

\subsection{Parametrizations of the Differential Cross Sections}

Several authors have investigated elastic hadron scattering by means
of parametrizations for the scattering amplitude and fits to the differential
cross section data. The extraction of the Profile, Eikonal and
Inelastic Overlap functions in the $b$-space and, in some special
cases, the Eikonal in the $q^2$-space, has led to important and novel 
results. For our purposes, some representative works are listed in
References \cite{kawa1,kawa2,fp}, 
\cite{franco}, \cite{myy}, \cite{sanivalin,as,fbv},
\cite{lw,chou,francahama,fms,bcsw,fearnley,lombard,franca,kppp}
and will be discussed in what follows. We first recall that
typical extracted results concern geometrical aspects (radius, central
opacity), differences between charge distributions and hadronic matter
distributions, existence or not of eikonal zeros in the $q^2$-space and,
more recently, connections with pomerons, reggeons and nonperturbative
QCD aspects.

The basic input in all these analyses is the parametrization of the
scattering amplitude as a sum of exponentials in $q^2$ 
(\cite{kawa1,kawa2,myy} are exceptions) and fits
to the differential cross section data. This parametrization allows
analytical expressions for the Fourier transform of the amplitude,
providing also analytical expressions for the quantities of interest
in the $b$-space. However, three kind of problems put serious limitations
on the information that may be extracted through this procedure:

(1) Experimental data are available only over finite regions
of the momentum transfer (which in general is small, 
$q^2 < $ 6 GeV$^2$) and the Fourier transform demands integration
from $q^2$ = 0 to infinity. This means that any fit is biased by 
extrapolations, a problem that has been well put by R. Lombard
\cite{lombard} : ``...extrapolating the measured differential cross
section can be done in an infinite number of manners. Some extrapolated
curves may look unphysical, but they can not be excluded on
mathematical grounds."

(2) The exponential parametrization allows analytical determination
of the quantities in the $b$-space and also the statistical
uncertainties, by means of error propagation from the fit parameters. However,
in this case, the translation to the $q^2$-space, Eq. (5), can not be 
analytically
performed and neither the error propagation (through standard procedures).
As a consequence, the unavoidable uncertainties from the fit extrapolations
can not, in principle, be taken into account.

(3) Some analyses introduce additional constraints in the
fit parameters so as to test some theoretical ideas or to obtain the
dependence of some parameters as function of the energy. Although, in
general, the number of free parameters is smaller than in ``empirical"
analyses, this procedure certainly introduces theoretical bias and 
can not be considered model-independent.

All the analyses based on parametrizations of the differential cross
sections face with one or more of the above mentioned problems.
Only to exemplify these aspects, we recall that Lombard and
Wilkin treated the experimental data by generating an ensemble of fits
with acceptable $\chi^2$ and defining an averaged value for the eikonal 
within one standard deviation. Although the extrapolations 
had been  taken into account,
the limitations in the interval of momentum transfer with available data
did not allow to extract information on the eikonal at large or even
intermediated values of the momentum transfer \cite{lw}. In the
analysis by Amaldi and Schubert, statistical and scale errors from the
experimental data on $pp$ scattering have been propagated by means of a
numerical method, but only to the impact parameter space \cite{as}. That
is also the case in the work by Fearnley, who treated both $pp$ and
$\bar{p}p$ scattering \cite{fearnley}. In the analysis by Sanielevici
and Valin \cite{sanivalin} and later by Furget, Buenerd and Valin 
\cite{fbv}, the eikonal has been determined in the $q^2$-space,
but without uncertainties from the fit parameters. At last, we recall that
the parametrization by Amaldi and Schubert (also used in \cite{fbv}) is
strongly model dependent, since the only parameter depending on the
energy is constrained by \cite{as}

\begin{eqnarray}
\alpha(s) = 
\left[\frac{\sigma_{tot}(s)}{\sigma_{tot}(\sqrt s = 23.5\ 
\textrm{GeV})}\right]
 [1 - i\rho(s)]. \nonumber
\end{eqnarray}
This embodies the geometrical scaling hypothesis, which is
well known to be violated at the Collider energy ($\sqrt s =$ 546 GeV).

\subsection{Strategies and ensembles}

Based on the above discussion, we have developed a model independent analysis
aimed to optimize some aspects of this kind of approach and, simultaneously,
to establish the confidence intervals of the extracted informations on
well based statistical grounds. The main points are the following:

(1) Compilation of the widest amount of experimental information 
presently available.

(2) Choice of a parametrization and a fit procedure as model
independent as possible.

(3) Development of a statistical procedure in order to estimate
the eikonal uncertainties in the momentum transfer space.

(4) To perform a systematic investigation of the effect in the
extracted eikonal, related with the existence or not of 
experimental data at large values of the momentum transfer.

To this end, we first make use of the empirical evidence that at 
$\sqrt s \sim $ 20 - 60 GeV and for momentum transfer above $q^2 \sim$ 
3 GeV$^2$,
the $pp$ differential cross section data are almost energy independent
\cite{donnachielandshoff,faissler}. This is illustrated in Fig. 1
for $pp$ scattering at 19.5 GeV $\leq \sqrt s \leq$ 62.5 GeV. The data
show agreement with a scattering amplitude parametrized by a sum of two
exponentials in $q^2$ and a fit through the CERN-MINUIT routine \cite{minuit}
has given

\begin{eqnarray}
F(q) = 0.032 e^{- 1.06 q^2} + 0.0012 e^{- 0.42 q^2}, 
\end{eqnarray}
with $\chi^2$/DOF = 294/162 $\sim$ 1.8. The result of the fit and the
contributions from the above two 
components are  displayed in Fig. 1. This suggests that the data at
$\sqrt s =$ 27.5 GeV, covering the region
$5.5 \leq q^2 \leq 14.2$ GeV may be included in the analyses of individual sets 
at other energies and our first point is to investigate how far can we go by
including these data in different sets for $pp$ and $\bar{p}p$ scattering.

\begin{figure}
\resizebox{0.48\textwidth}{!}{\includegraphics{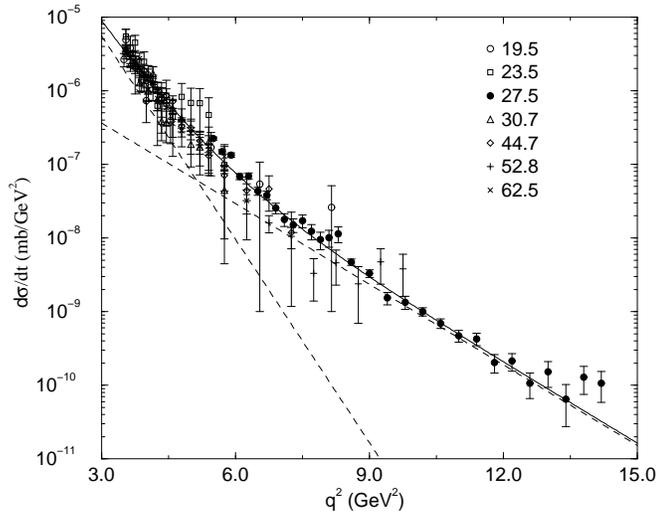}}
\caption{Differential cross section data for $pp$ scattering
above $q^2 =$ 3.0 GeV and fit through Eq. (13), with the
contributions from the two exponential terms.}
\label{fig:1}       
\end{figure}

Our strategy is to consider two different ensembles of data, initially 
characterized
and denoted as follows.

\begin{description}
\item[Ensemble A]: Original sets of data at each energy: 7 sets for $pp$
scattering and 7 sets for $\bar{p}p$ scattering (Sec. 3.1).
\item[Ensemble B]: Sets of ensemble A including in each one the data at
$\sqrt s =$ 27.5 GeV.
\end{description}

Once selected a parametrization for the scattering amplitude (to be discussed
in what follows) the validity or not of ensemble B 
(that is, the compatibility or not of the data at 27.5 GeV with the
original set)
shall be checked by means of
fits through the CERN-MINUIT routine \cite{minuit} and standard statistical
interpretation of the fit results \cite{bevington}. 

We note that the addition of the
data at $\sqrt s$ = 27.5 GeV to $pp$ data at 52.8 GeV has been previously
 explored by
Sanielevici and Valin \cite{sanivalin}. The novel aspect here is to perform
 a systematic 
investigation on the validity of this assumption in other energies and in
$\bar{p}p$ scattering.

\section{Parametrization, fitting of the data and results}
\label{sec:4}

\subsection{Parametrization and fit procedure}

We consider the standard parametrization of the scattering amplitude
as a sum of exponentials in $q^2$:

\begin{eqnarray}
 F(s,q) \quad \rightarrow \quad  \sum_{i=1}^{n} \alpha_{i}
 e^{-\beta_{i} q^{2}}. \nonumber
\end{eqnarray}
The specific structure of the parametrization was determined 
by means of the following model-independent procedure.

We have first started with the $pp$ scattering at $\sqrt s =$ 52.8 GeV,
since the original set (ensemble A) corresponds to the highest interval
in the momentum transfer with available data, namely  
0.01 $ < q^2 \leq $ 9.8 GeV$^2$ and the optical point, Eq. (12). 
 The diffraction
peak region, 0.01 $< q^2 \leq$ 0.5 GeV$^2$, is characterized by a change
of the slope around $q^2 \sim $ 0.13 GeV$^2$ \cite{castsangui} and the
dominance of the imaginary part of the scattering amplitude, since
$\rho =$ 0.078 \cite{as}. For these reasons we have represented the imaginary
part of the amplitude by a sum of two exponentials. From the visual 
examination of
the data, we inferred initial values for the free parameters and then, 
they have been
statistically determined by means of the CERN-MINUIT routine. Similar
procedure has been used in the region of large momentum transfer
(3.0 $< q^2 \leq 9.8$ GeV$^2$), leading to the determination of two 
additional and independent exponential terms. In order to generate the
diffraction minimum we added to the previous fixed results an exponential
term with negative sign. Looking for the simplest approach, with
an economical number of free parameters, we have tested the possibility
that two exponential terms could represent the real part of the
amplitude. The only constraint corresponds to the definition of the
$\rho$ parameter, Eq. (8). With the ``experimental" $\rho$ value as
input and using as initial values for the free parameters those
determined previously, the fit led to a statistically consistent final 
result, $\chi^2$/DOF = 323/196 = 1.65.

From this result at $\sqrt s$ = 52.8 GeV we fitted the $pp$ data in a
sequence of nearest
energies, using as initial values for the free parameters those previously
obtained in each case. The same procedure has been applied to $\bar{p}p$ data
beginning at $\sqrt s$ = 53 GeV and then going through the sequence of 
nearest energies.
The values of the fit parameters with ensemble A have also been used
as initial values in the fits with ensemble B. 

With this procedure we have obtained a good reproduction of the experimental
data with the following par\-ametriza\-tions for the real and imaginary
 parts of 
the scattering amplitude:

\begin{eqnarray}
\mathrm{Re}\ F(s,q) = \mu \sum_{j=1}^{2} \alpha_{j} e^{-\beta_{j} q^{2}},
\end{eqnarray}

\begin{eqnarray}
\mathrm{Im}\ F(s,q) =  \sum_{j=1}^{n} \alpha_{j} e^{-\beta_{j} q^{2}}, 
\end{eqnarray}
where

\begin{eqnarray}
\mu = {\rho(s) \over \alpha_{1} + \alpha_{2}} \sum_{j=1}^{n} \alpha_{j},
\end{eqnarray}
$\alpha_{j}, \beta_{j}$ (j=1,2,...,n), are real free parameters and $\rho(s)$ is 
the  value extracted from experiments at each energy. The
number of exponentials terms depends on the reaction and data set as shown
in what follows.

It should be noted that, with this procedure, the fit parameters are
completely free, since, here,  we are not interested in obtaining 
dependences on the energy and/or reaction. Our aim is the best statistical
result, without any constraint in the fit parameters.

\subsection{Fitting results}

The central values of the free parameters in all the fits presenting
statistical consistence are displayed in Tables 1 - 3, together with other
statistical information. The errors in the parameters 
(not displayed) correspond to an
increase of the $\chi^2$ by one unity and involve variances and 
covariances which will be discussed in the next Subsection.

The results from $pp$ scattering with ensembles A and B appear in Tables
1 and 2. We have found that the data at $\sqrt s =$ 13.8 GeV are
not compatible with the data at 27.5 GeV (ensemble B). In the case of 
$\bar{p}p$ scattering
none of the data sets are compatible with the $pp$ data at
27.5 GeV (ensemble B) and the results with ensemble A are 
displayed in Table 3. Therefore, in what follows, ensemble B (data at
$\sqrt s =$ 27.5 GeV added) corresponds only to $pp$ scattering
at 6 energies: 19.5, 23.5, 30.7, 44.7, 52.8 and 62.5 GeV.
The fit results for $pp$ scattering together with the 
experimental data
in ensembles A and B are shown in Figs. 2 and 3 and those for $\bar{p}p$
scattering with ensemble A in Fig. 4.

\begin{table*}
\begin{center}
\caption{Fitting results at each energy (in GeV) for $pp$ scattering with 
{\it Ensemble} A:
values of the free parameters in GeV$^{-2}$, experimental $\rho$ value, 
maximum value of the momentum transfer in GeV$^2$, number 
of experimental points (N) and  $\protect\chi^2$ per degree of freedom.} 
\label{tab:1}
\begin{tabular}{lllllllll}
\hline\noalign{\smallskip}
$\sqrt s$:  &13.8&19.5&23.5 & 30.7 & 44.7 & 52.8 & 62.5 \\
\noalign{\smallskip}\hline\noalign{\smallskip}
$\alpha_{1}$&$-5.698 \times 10^{-4}$&$-1.465 \times 10^{-2}$& 
$-0.2390$ & $-4.115 \times 10^{-2}$ & $-1.090 \times 10^{-2}$ & $-2.123 
\times 10^{-2}$ & $-4.281 \times 10^{-2}$ \\
$\alpha_{2}$&$8.775 \times 10^{-4}$&0.3155 & 3.267 & 3.569 & 0.6257 & 1.147 
& 2.349\\
$\alpha_{3}$&5.983&4.180&0.2299 & - & 3.672 & 3.662 & 0.1802 \\
$\alpha_{4}$&-3.670&-3.106 & - & - & $-3.062$ & $-3.070$ & - \\
$\alpha_{5}$&5.514&6.588& 4.657 & 4.641 & 7.432 & 7.020 & 6.326\\
$\beta_{1}$&173.2&0.7601& 1.141 & 0.9121&  0.6919 & 0.7983 & 0.9444 \\
$\beta_{2}$&3.711&6.085&8.565&8.303&31.77&17.39& 11.13\\
$\beta_{3}$&2.796&2.341& 1.285 & - & 2.172& 2.277 & 2.832\\
$\beta_{4}$&2.425&2.165 & - & - & 2.050 &2.165& - \\	
$\beta_{5}$&6.428&6.083&4.279&4.253 &6.094&5.739& 5.140 \\
$\rho$&-0.074&0.019&0.02&0.042 &0.062&0.078&0.095\\
$q^2_{\mathrm{max}}$&2.82&8.15&5.75&5.75&7.25&9.75&6.25\\
N&100&123&134&173&208&206&125\\
$\chi^{2}$/DOF&2.03&2.96&1.14&1.00&2.13&1.65&1.16\\
\noalign{\smallskip}\hline
\end{tabular}
\end{center}
\end{table*}

\begin{table*}
\begin{center}
\caption{Fitting results for $pp$ scattering with {\it Ensemble} B:
same legend as in Table \ref{tab:1}.}
\label{tab:2}
\begin{tabular}{llllllll}
\hline\noalign{\smallskip}
$\sqrt s$:  &19.5&23.5 & 30.7 & 44.7 & 52.8 & 62.5 \\
\hline
$\alpha_{1}$&$-9.012 \times 10^{-3}$& $-0.1935$ & $-4.044 
\times 10^{-2}$ & $-1.014 \times 10^{-2}$ & $-2.683 
\times 10^{-2}$ & $-4.238 \times 10^{-2}$ \\ 
$\alpha_{2}$&0.2317 & 3.619 & 3.474 & 0.6030 & 1.198 & 2.023\\
$\alpha_{3}$&4.192 & $0.1538$ & - & 3.699 & 3.652 & 0.3443 \\ 
$\alpha_{4}$&-3.101 & - & - & $-3.039$ & $-3.078$ &$9.109 
\times 10^{-2}$\\
$\alpha_{5}$&6.650& 4.326 & 4.742 & 7.404 & 7.002 & 6.405\\
$\alpha_{6}$ &$-1.572 \times 10^{-5}$& $-9.176 
\times 10^{-4}$ &$-1.968 \times 10^{-3}$& $-1.256 \times 10^{-4}$ &$-1.673 
\times 10^{-3}$& $-3.263 \times 10^{-3}$\\
$\beta_{1}$&0.6177& 0.9112 & 0.9771 & 0.6685 
& 0.9369 &1.062 \\
$\beta_{2}$&6.055&8.178&8.424&32.62&16.94&11.96\\
$\beta_{3}$&2.351& 0.9123 &-& 2.227& 2.272 &3.352\\
$\beta_{4}$&2.167 &-& - &2.095 &2.169 &3.335\\
$\beta_{5}$&6.094&4.148&4.275 &6.135&5.704&5.290 \\
$\beta_{6}$&$6.323 \times 10^{-2}$ & 0.3558&
0.4210&0.2113&0.4061 &0.4671\\
$\rho$&0.019&0.02&0.042&0.062&0.078&0.095\\
$q^2_{\mathrm{max}}$&14.2&14.2&14.2&14.2&14.2&14.2\\
N&153&164&203&238&236&155\\
$\chi^{2}$/DOF&2.80&1.20&1.28&2.13&2.07&1.51\\
\noalign{\smallskip}\hline
\end{tabular}
\end{center}
\end{table*}

\begin{table*}
\begin{center}
\caption{Fitting results for $\bar{p}p$ scattering with {\it Ensemble} A:
same legend as in Table \protect\ref{tab:1}. }
\label{tab:3}
\begin{tabular}{lllllllll}
\hline\noalign{\smallskip}
$\sqrt s$: &14 & 19 & 31 & 52.8 &62.5 & 546  & 1800\\
\hline
$\alpha_{1}$& $-1.423$ & $-3.627 \times 10^{-2}$ & - & 
$-3.038 \times 10^{-2}$ & - & $-0.2038$ & - \\
$\alpha_{2}$ & 7.247 & 7.525 & - & 0.9131 & - & 2.854  & -\\
$\alpha_{3}$ & 2.768 & 0.9795 & - & 3.471 & - & 0.3705  & - \\
$\alpha_{4}$ & - & - & - & $-3.214$  & - & -  & -\\ 
$\alpha_{5}$& - & - & 8.601 & 7.787 & 7.377 & 9.953  & 16.60\\
$\alpha_{6}$& - & - & - & - & 1.659  & -  & - \\ 
$\beta_{1}$& 2.027 & 0.6502 & - & 0.9858 & - & 1.519  & - \\ 
$\beta_{2}$&6.429&6.320&-&151.8&-& 14.27 &-\\
$\beta_{3}$& 2.487 & 2.770 & - & 2.237 & - & 3.235  & - \\
$\beta_{4}$ & - & - & - &2.213 &-& - & - \\ 
$\beta_{5}$&-&-&5.845&5.990 & 5.457&6.470&8.371\\
$\beta_{6}$&-&-&-&- & 17.72& - &-\\
$\rho$&0.014&0.029&0.065&0.101&0.12&0.135&0.14\\
$q^2_{\mathrm{max}}$&2.45&4.45&0.85&3.52&0.85&1.53&0.626\\
N&61&22&23&52&24&122&47\\
$\chi^{2}$/DOF&1.00&0.57&1.53&1.85&0.72&1.01&0.92\\
\noalign{\smallskip}\hline
\end{tabular}
\end{center}
\end{table*}

 Typical contributions to the
differential cross section from the real and imaginary parts of the
scattering amplitude are illustrated in Figs. 5 and 6 for $pp$
scattering at 52.8  GeV and 30.7 GeV, respectively.
These ``contributions", generated by the fit procedure without model
dependence (Sec. 4.1), allow to infer some interesting ``empirical"
results. We see that both the real and imaginary parts of the
amplitude present only one zero (change of sign), the former at
small values of the momentum transfer and the later at the dip position.
Also, with the exception of the dip region, which is filled by the
real part, the imaginary part dominates at all values of the
momentum transfer.  It is also important to note that,
although not imposed in the fit procedure, the position of the zero of the
real part (ensemble B) is in agreement with a theorem recently demonstrated
by A. Martin, which states that the real part (even amplitude) changes sign at
$q^2$ above $\approx$ 0.1 GeV$^2$ \cite{martinzero}.

\begin{figure}
\resizebox{0.48\textwidth}{!}{\includegraphics{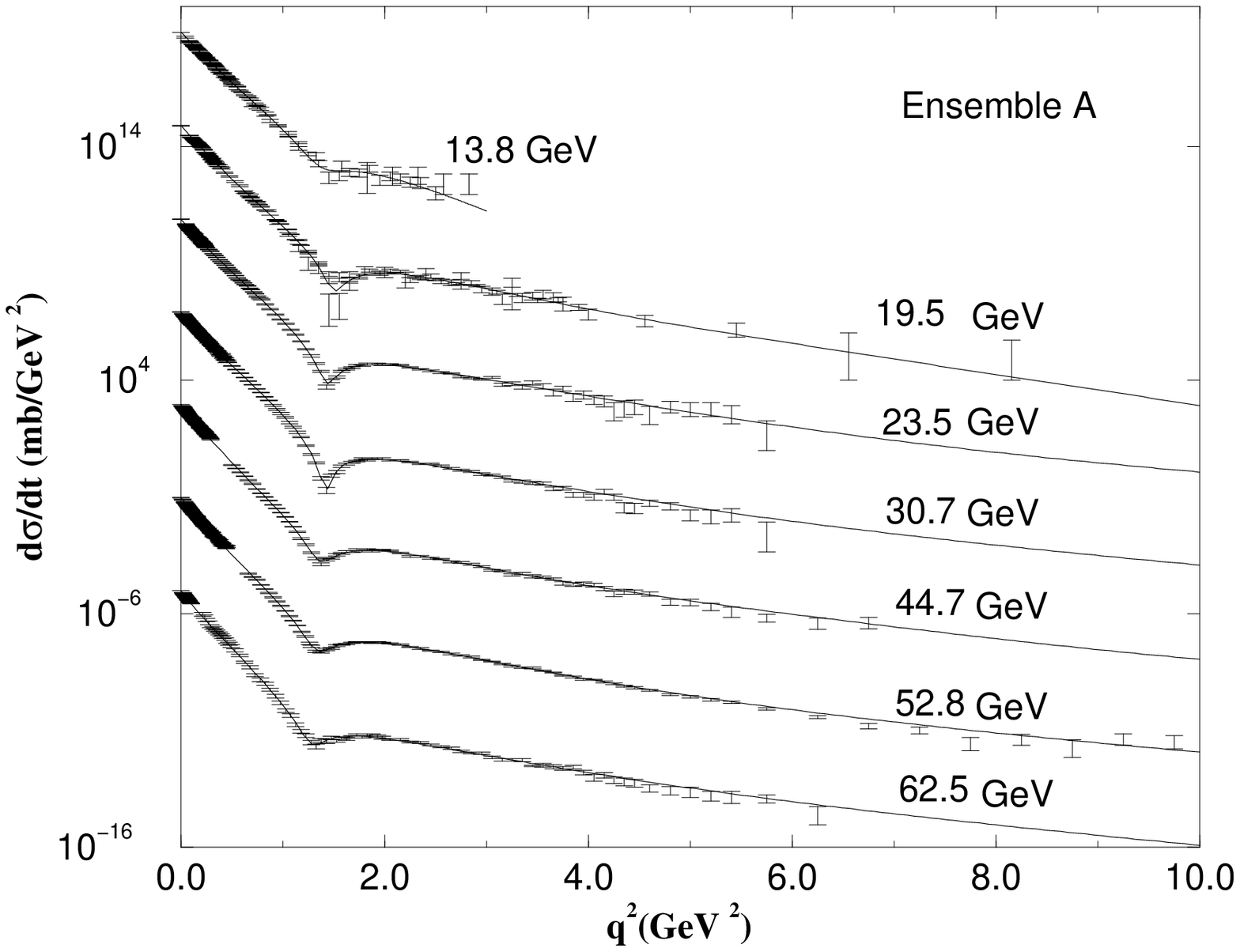}}
\resizebox{0.48\textwidth}{!}{\includegraphics{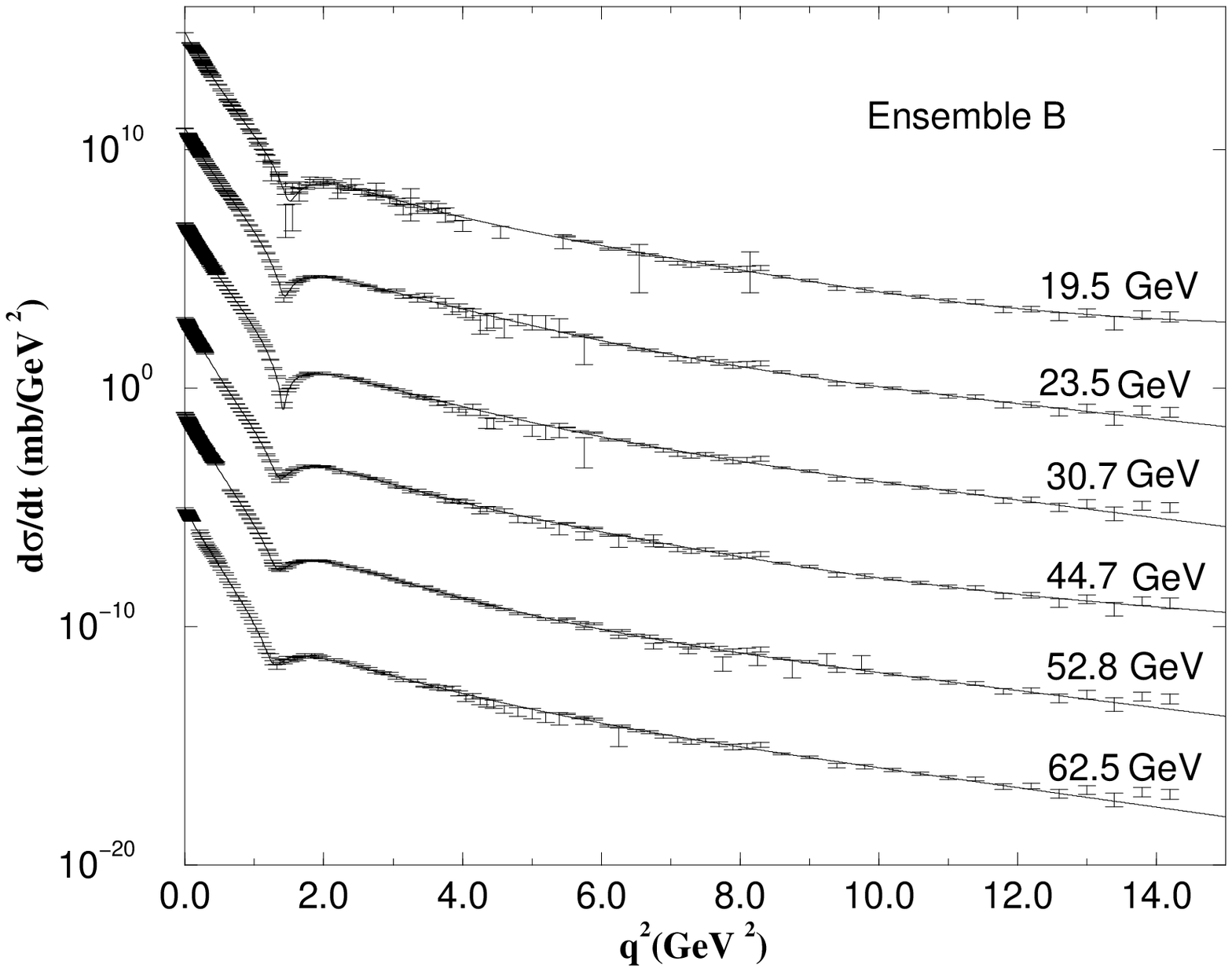}}
\caption{Fit results for $pp$ scattering with ensembles A and B.}
\label{fig:2}       
\end{figure}

\begin{figure}
\resizebox{0.48\textwidth}{!}{\includegraphics{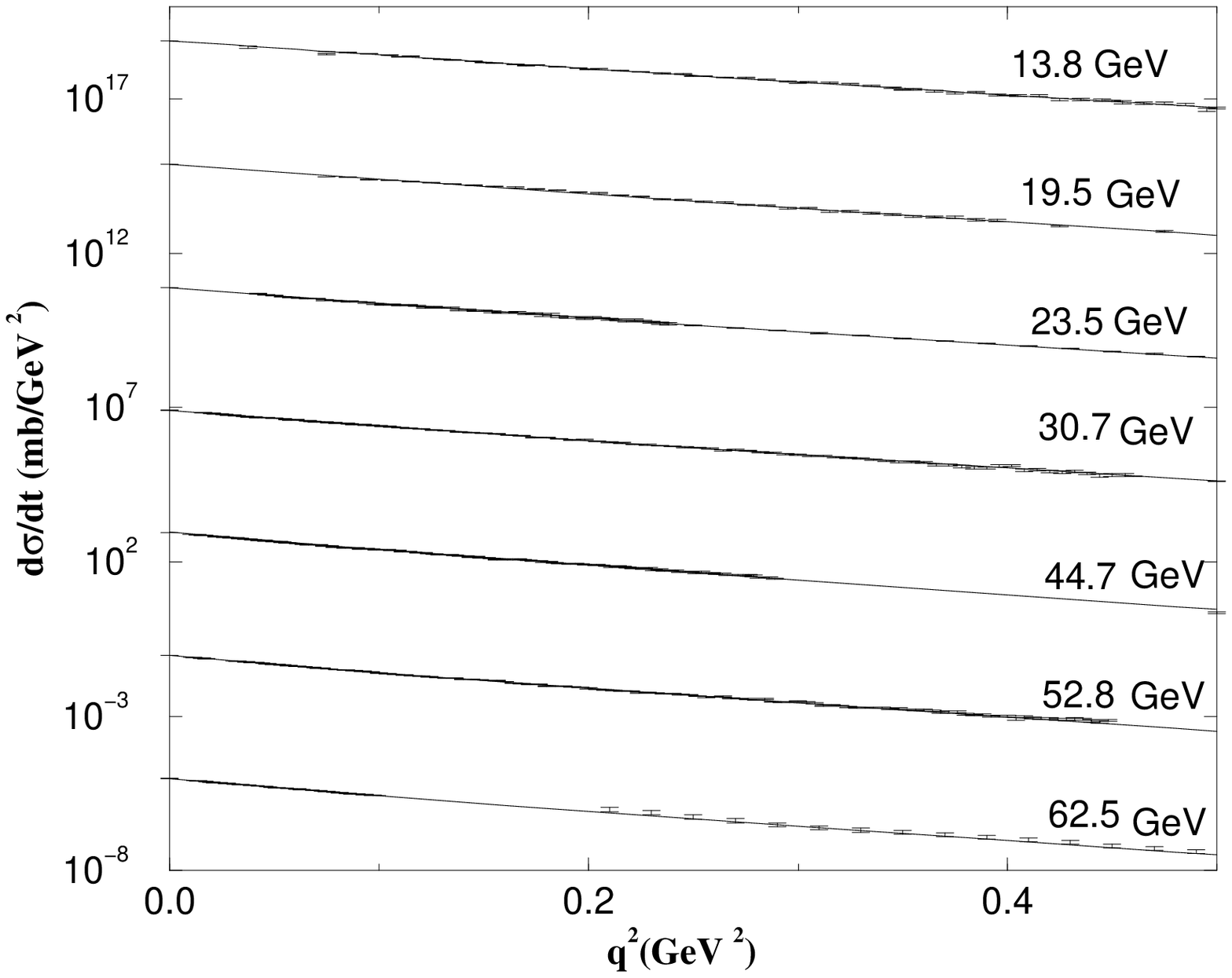}}
\resizebox{0.48\textwidth}{!}{\includegraphics{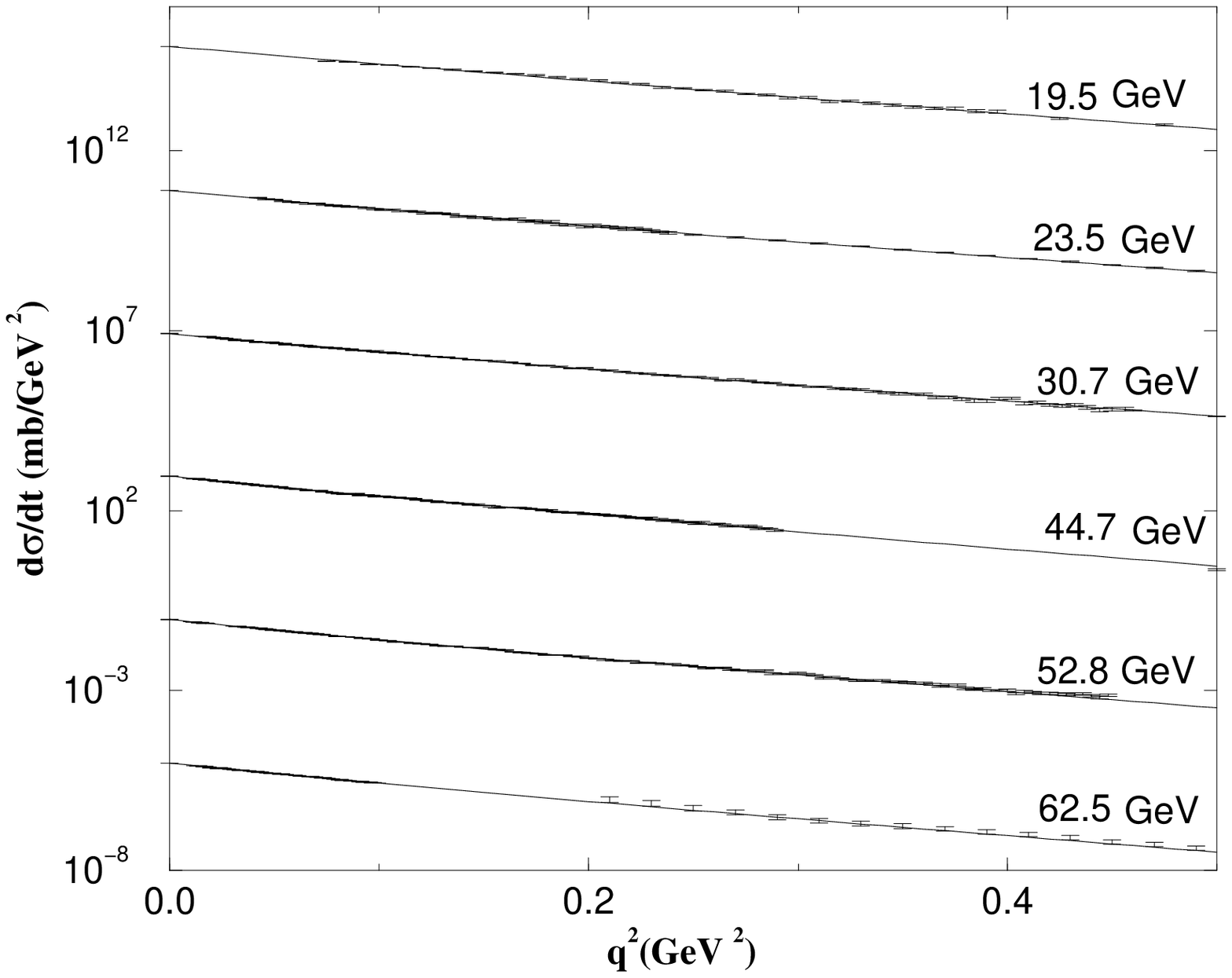}}
\caption{Same as Fig. 2 in the region of small momentum transfer.}
\label{fig:3}       
\end{figure}

\begin{figure}
\resizebox{0.48\textwidth}{!}{\includegraphics{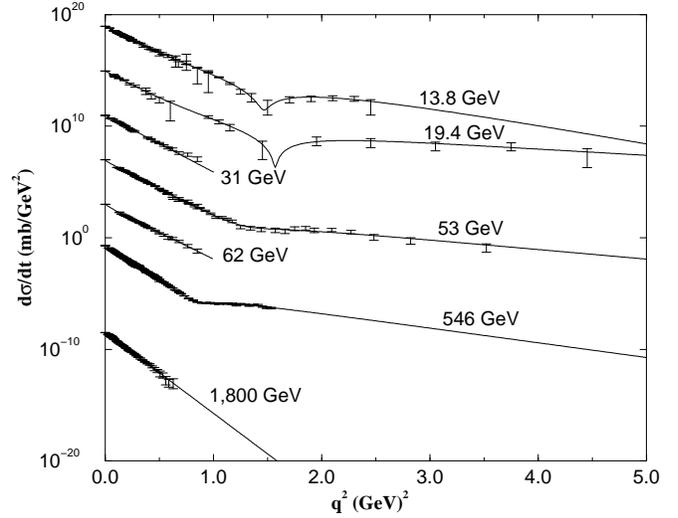}}
\caption{Fits results for $\bar{p}p$ scattering with ensemble A.}
\label{fig:4}       
\end{figure}

\begin{figure}
\resizebox{0.48\textwidth}{!}{\includegraphics{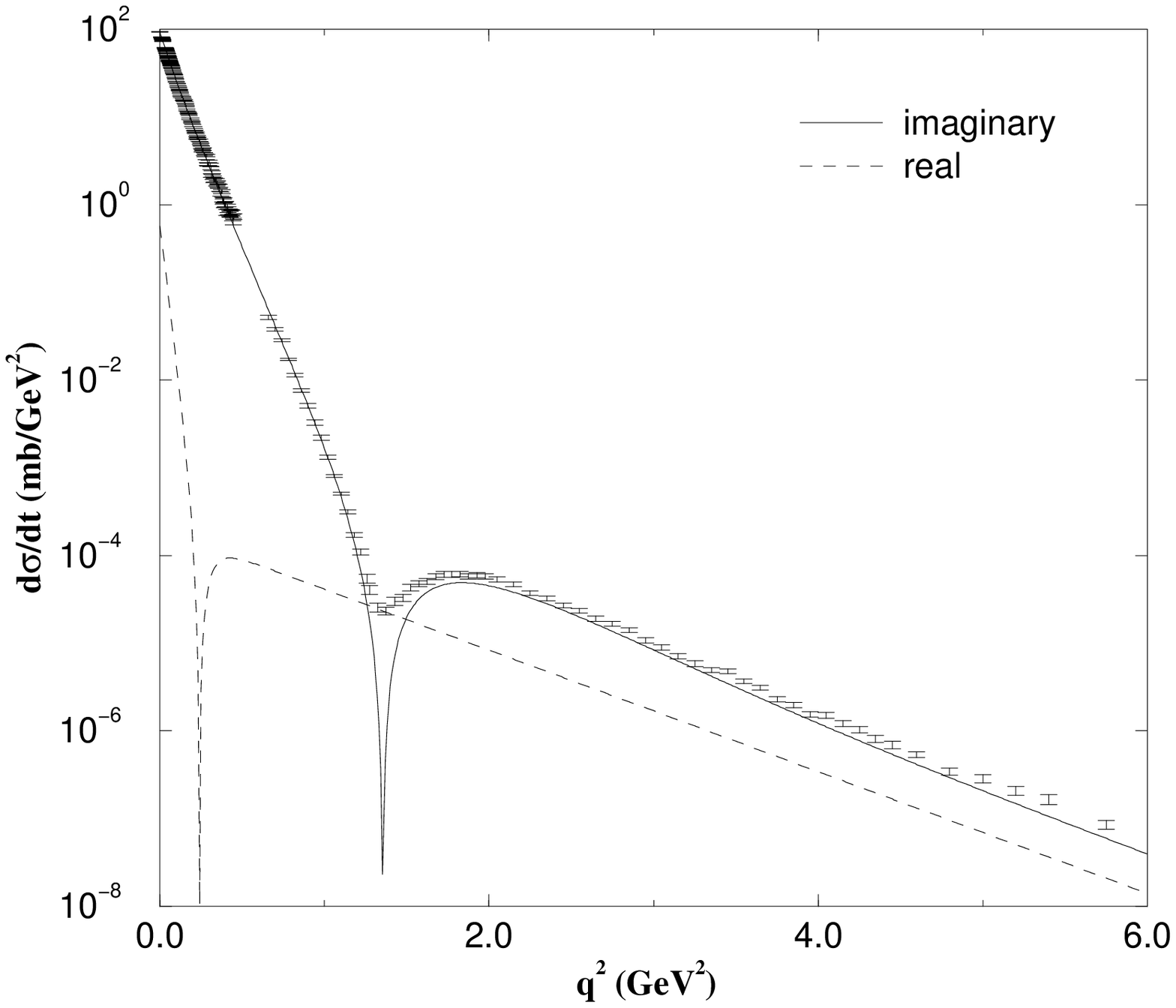}}
\resizebox{0.48\textwidth}{!}{\includegraphics{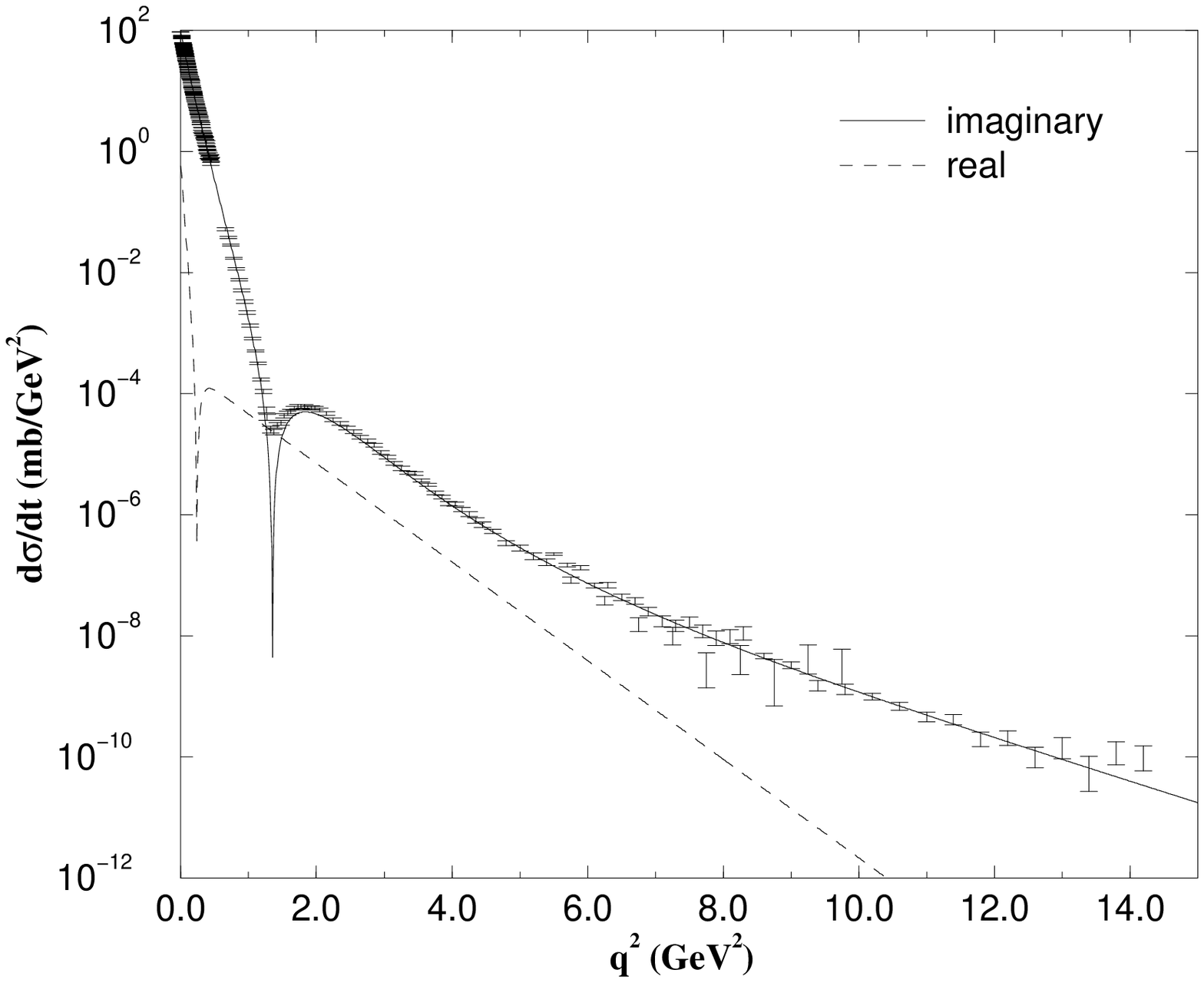}}
\caption{Proton-proton scattering at $\sqrt s =$ 52.8 GeV.
Contributions to the differential cross section from the real and 
imaginary parts of the scattering amplitude with ensembles A (up) 
and B (down).}
\label{fig:5}       
\end{figure}

\begin{figure}
\resizebox{0.48\textwidth}{!}{\includegraphics{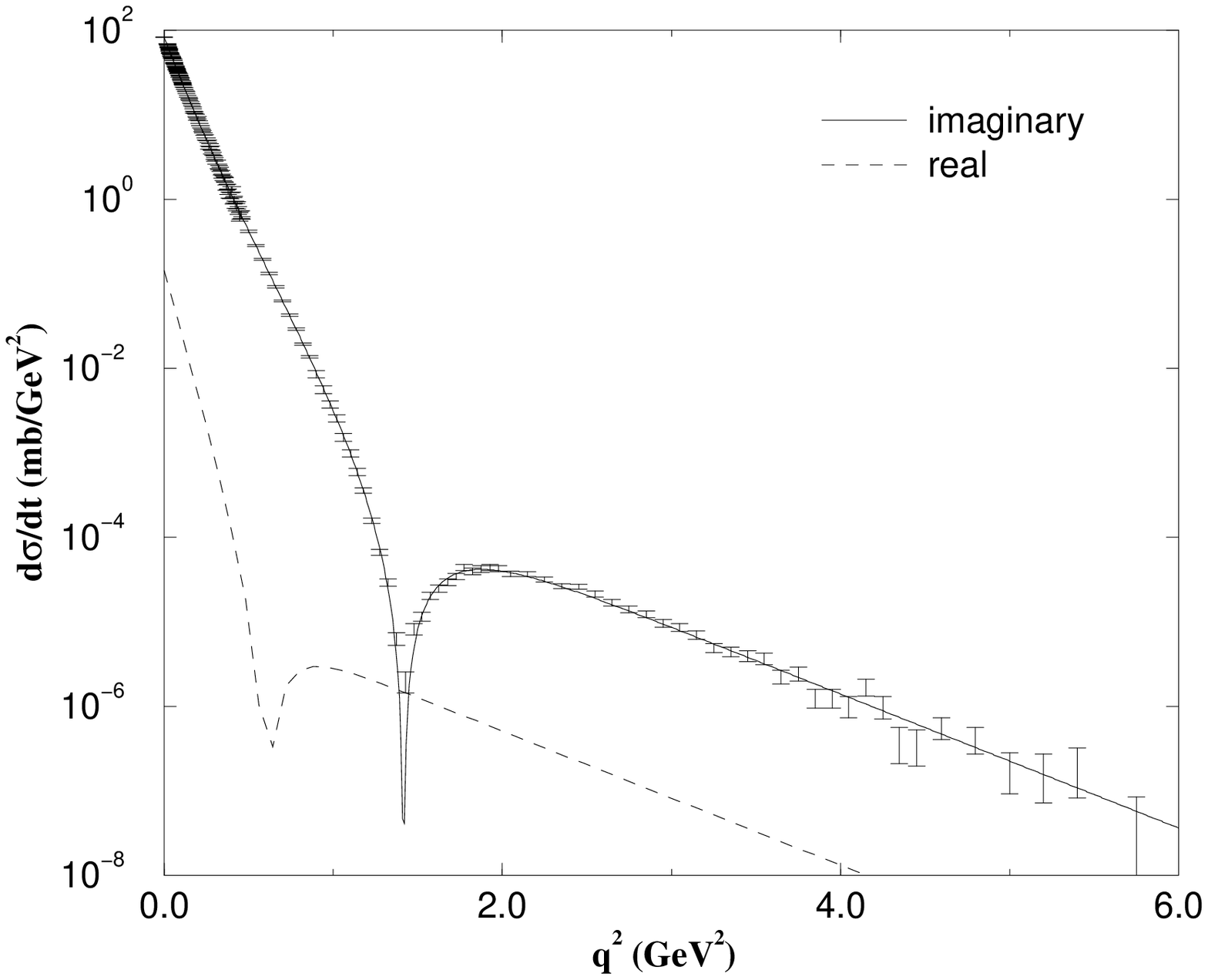}}
\resizebox{0.48\textwidth}{!}{\includegraphics{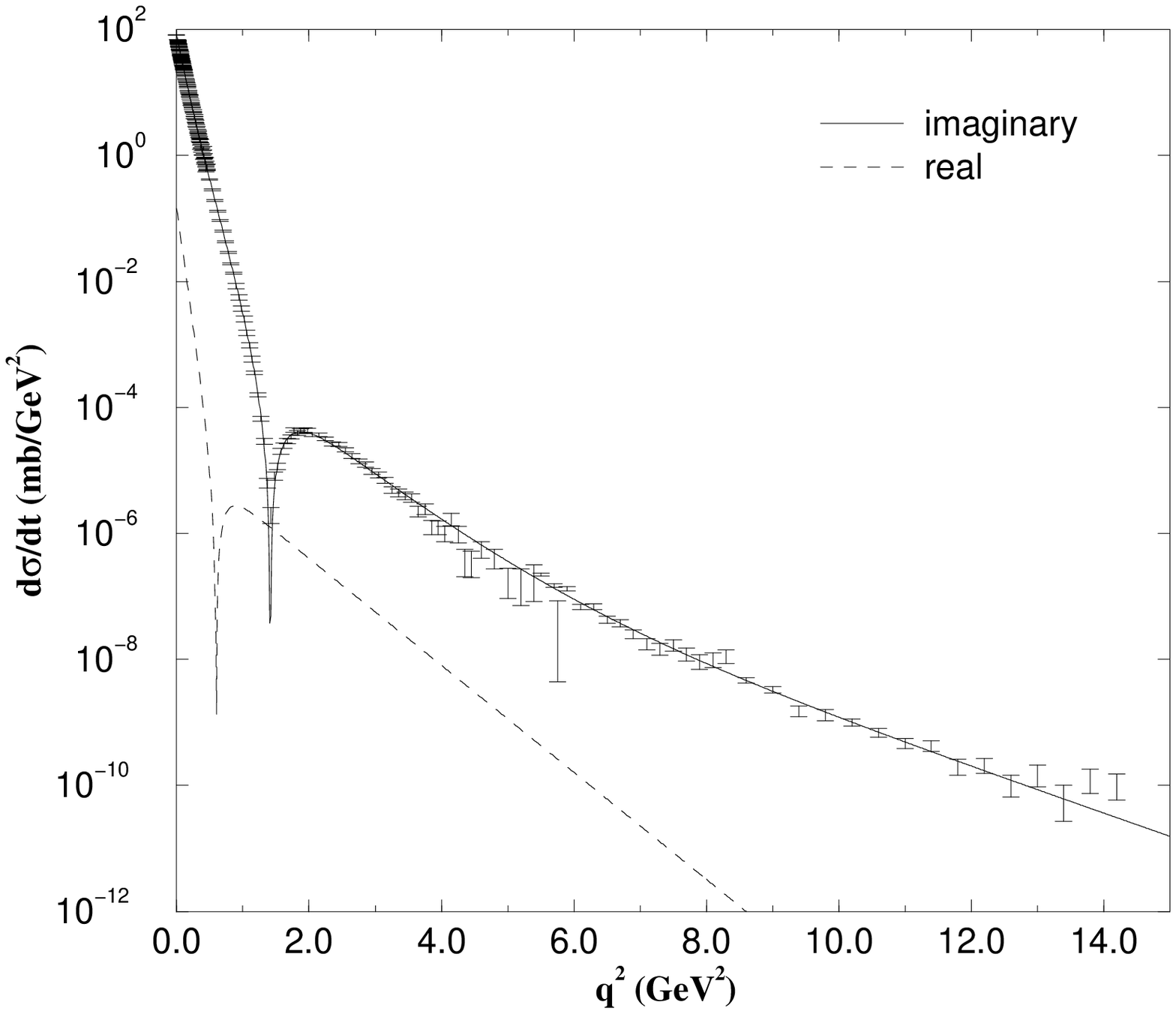}}
\caption{Same as Fig. 5 for proton-proton scattering at $\sqrt s =$ 30.7 GeV.}
\label{fig:6}       
\end{figure}

\subsection{Uncertainties and error propagation}

As mentioned before, we are interested in the determination of the confidence
interval of the extracted information, meanly related with the existence or not 
of 
experimental data at intermediate and large values of the momentum transfer.
For this reason we made use of the same parametrization and fit procedure for 
all the sets
in ensemble A, even in the cases where the available data concern only the 
diffraction peak,
as shown in Fig. 4. Certainly, this lack of information will be mirrored in the
associated confidence interval and that is the point we are interested in.

These considerations may be quantitatively exemplified as follows. In each fit, 
the error matrix provides the variances and covariances associated with 
each parameter
(the numerical values do not appear in Tables 1 - 3 due to lack of space,
but are available from the authors). By means of standard error propagation
\cite{bevington} the uncertainties in the free parameters, $\Delta \alpha_{j}$,
$\Delta \beta_{j}$, ($j$ = 1, 2, ...) have been propagated to the scattering
 amplitude,
Eqs. (14 - 16) and then to the differential cross section, Eq. (6), providing

\begin{eqnarray}
\frac{d\sigma}{dq^2} \pm \Delta \left(\frac{d\sigma}{dq^2}\right).
\end{eqnarray}
By adding and subtracting the corresponding uncertainties we may estimate
the confidence region associated with all the extrapolations, which cannot 
be excluded
on statistical grounds \cite{lombard}. A typical result with ensembles A and B
is illustrated in Fig. 7, for $pp$ scattering at $\sqrt s$= 23.5 GeV. We see
 that,
as expected, the effect
of adding the experimental data at $\sqrt s$= 27.5 GeV (when statistically
justified) is to reduce drastically the uncertainty region. That result will
be fundamental in the extraction of the empirical information on
the eikonal, as shown in what follows.

\begin{figure}
\resizebox{0.48\textwidth}{!}{\includegraphics{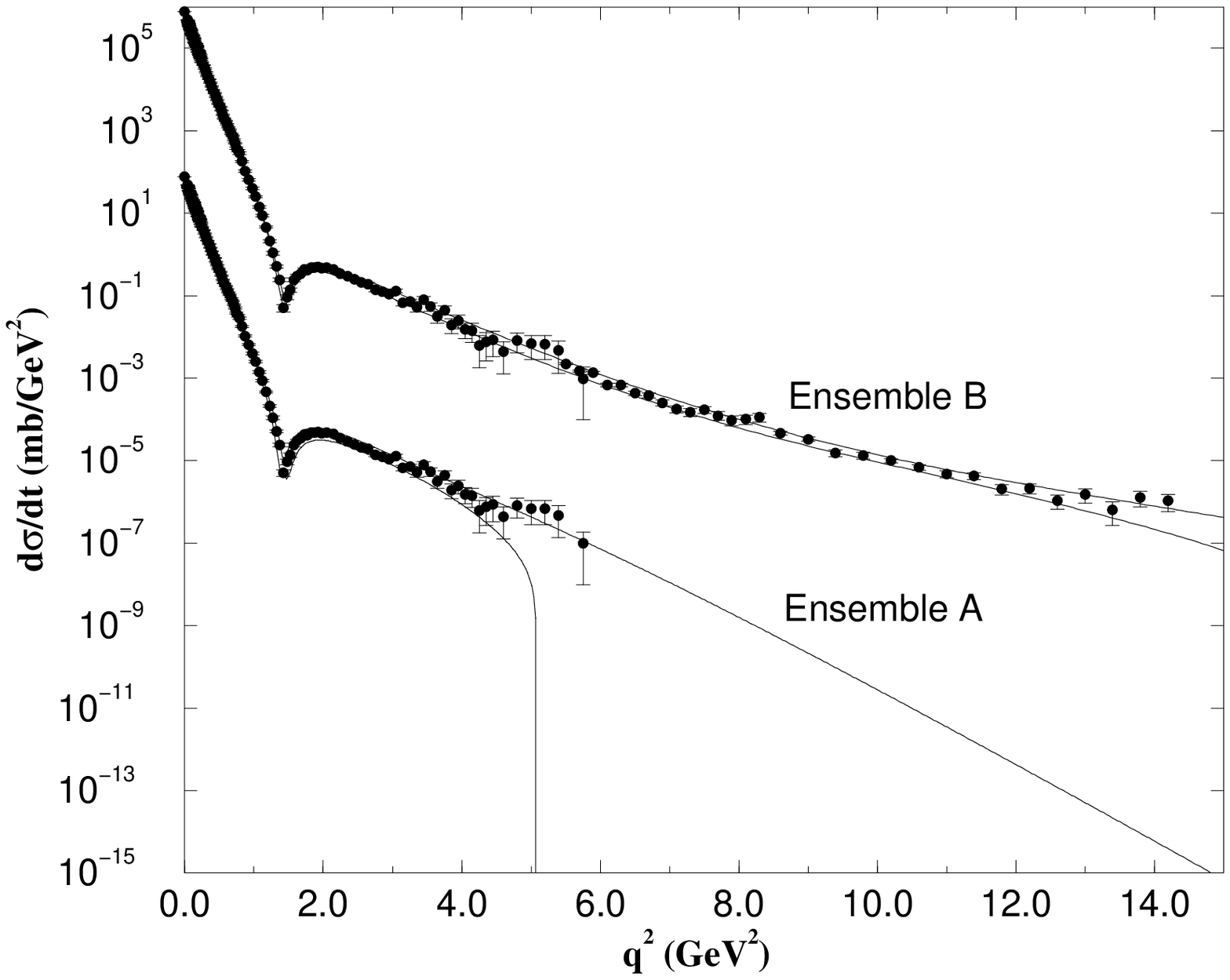}}
\caption{Elastic $pp$ scattering at $\sqrt s =$ 23.5 GeV.
Extrapolations from the fits with ensembles A and B.}
\label{fig:7}       
\end{figure}

Our fit results provide useful information for studies in the impact parameter
space, namely $\Gamma(s,b)$ and $G_{inel}(s,b)$. However, since we are 
interested here only in the eikonal in the momentum transfer space, we postpone 
the impact parameter analysis for a future work.

\section{Eikonal in the momentum transfer space}
\label{sec:5}

\subsection{Semi-analytical and numerical methods}

The first steps in going to $\chi(s,q)$ are the determination
of $\Gamma(s,b)$ and then $\tilde\chi(s,b)$. Denoting for short the real and 
imaginary parts by the subscripts $R$ and $I$, respectively, and inverting
Eq. (2) we have

\begin{eqnarray}
\tilde\chi_{R}(s,b)=\tan^{-1} \Big\{ \frac{\Gamma_{I}(s,b)}{ \Gamma_{R}(s,b) - 1} 
\Big\},
\end{eqnarray}

\begin{eqnarray}
\tilde\chi_{I}(s,b)= \ln \Big\{ \frac{1}{\sqrt{\Gamma^{2}_{I}(s,b) +
[1-\Gamma_{R}(s,b)]^{2}}} \Big\}.
\end{eqnarray}

By means of the Fourier transform, Eq. (1), the param\-etri\-za\-tion
(14 - 16) provide analytical expressions for $\Gamma_{R}(s,b)$,
$\Gamma_{I}(s,b)$, $\tilde\chi_{R}(s,b)$ and $\tilde\chi_{I}(s,b)$. Taking
into account the variances and covariances in the fit parameters,
error propagation gives the uncertainties for all these quantities.
An example is shown in Fig. 8, for the case of $pp$
scattering at $\sqrt s$ = 52.8 GeV (ensemble A).

\begin{figure}
\resizebox{0.48\textwidth}{!}{\includegraphics{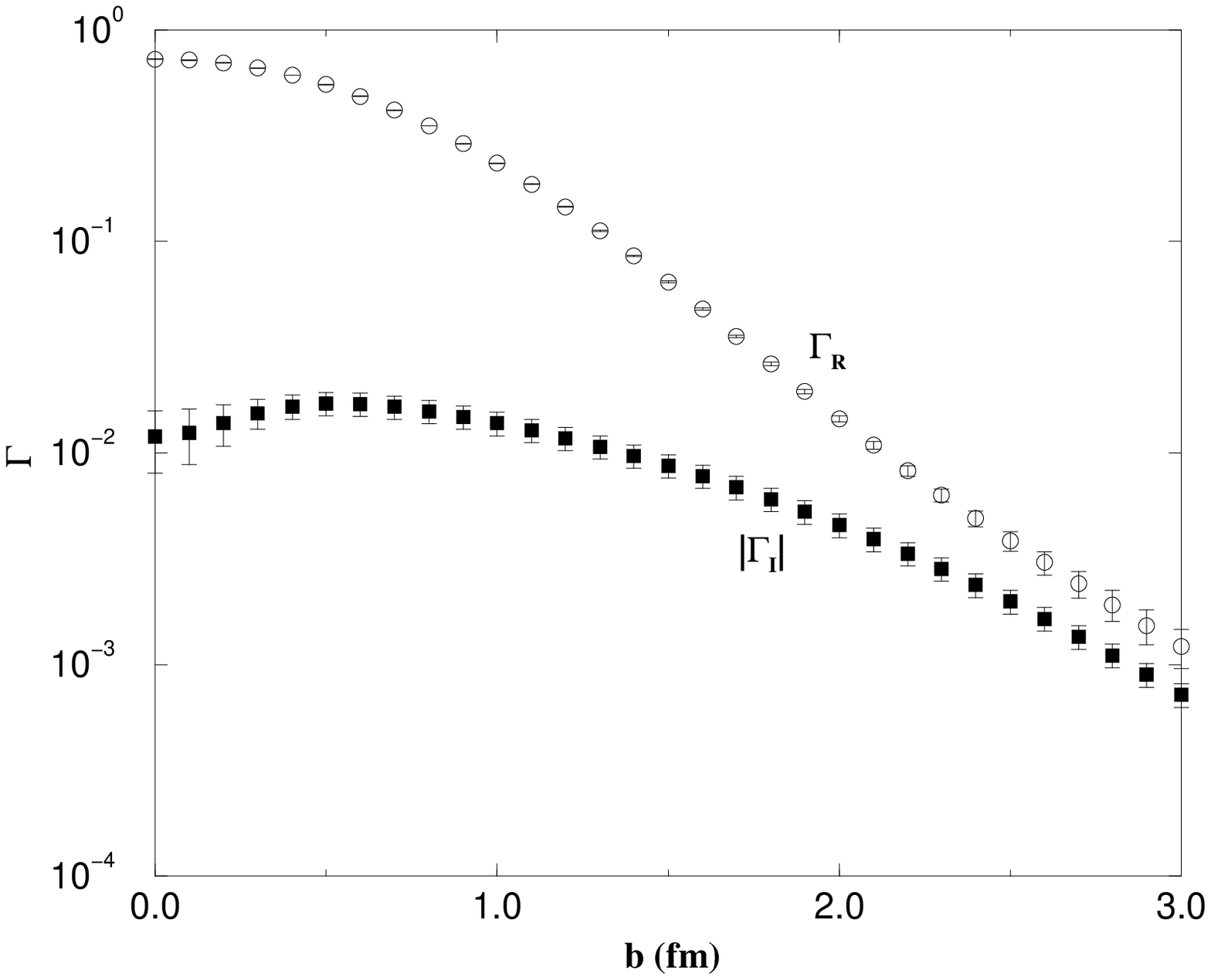}}
\resizebox{0.48\textwidth}{!}{\includegraphics{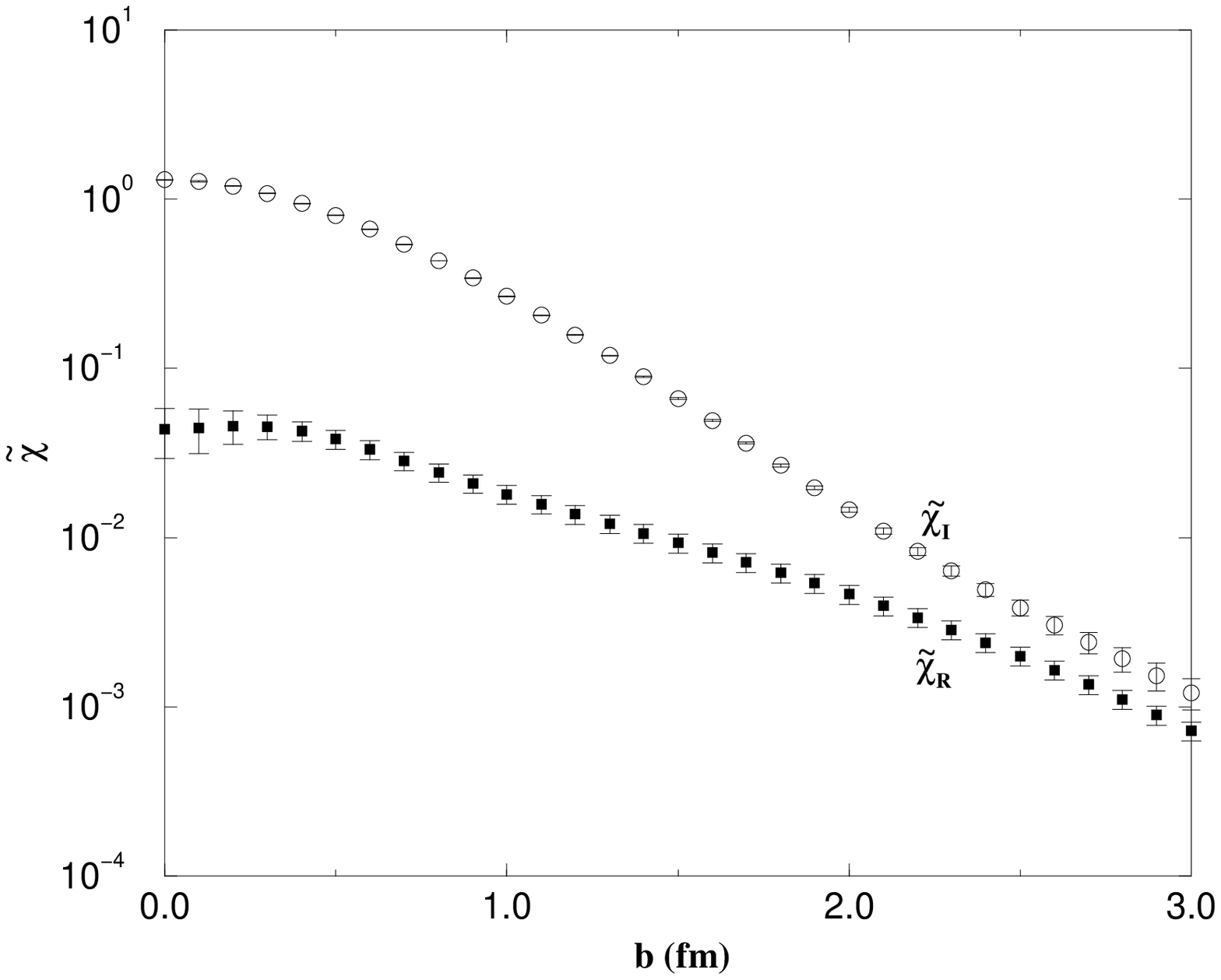}}
\caption{Real and imaginary  parts of the
Profile and
Eikonal functions for $pp$ at $\sqrt s =$ 52.8 GeV,
with ensemble A.}
\label{fig:8}       
\end{figure}

As shown in Sec. 2, the \textit{imaginary part of the Eikonal}
is connected with the Inelastic Overlap Function and the unitarity condition, 
Eqs. (2 - 4). It is usually named \textit{Opacity Function} and that is
the quantity we are interested to investigate. From the fit results,
together with error propagation, we have found that (see Fig. 8)

\begin{eqnarray}
\frac{\Gamma_{I}^{2}(s,b)}{[1-\Gamma_{R}(s,b)]^{2}} \ll1,
\nonumber
\end{eqnarray}
and therefore, the Opacity Function may be approximated by

\begin{eqnarray}
\tilde\chi_{I}(s,b)\approx\ln{1\over1-\Gamma_{R}(s,b)}
\end{eqnarray}
and the uncertainty $\Delta\chi_{I}$ determined directly from $\Delta\Gamma_{R}$
through error propagation.

The last step is to obtain the Fourier transform (5), which, due to the
structure of our parametrization, can not be analytically performed.
At this point we first make use of a numerical integration through the
NAG routine \cite{nag} and the results will be presented and discussed
later. Since the numerical integration does not allow standard error
propagation, we have developed the following approach, which we shall
name \textit{semi-analytical method}.

Generically, we can expand Eq. (19) in the form

\begin{equation}
\tilde\chi_{I}(s,b)=\Gamma_{R}(s,b)+R(s,b),
\label{expansao}
\end{equation}
where $R(s,b)$ corresponds to the remainder of the series. Performing 
the Fourier transform we obtain

\begin{equation}
\chi_{I}(s,q)=F_{I}(s,q)+R(s,q).
\label{expansaoq}
\end{equation}
Since the amplitude $F_{I}(s,q)$ and errors $\Delta F_{I}(s,q)$ are 
directly given by the fits, our task concerns the evaluation of 

\begin{equation}
R(s,q)= \int_{0}^{\infty} b db J_0(qb) R(s,b),
\label{Dtransf}
\end{equation}
with the corresponding errors, $\Delta R(s,q)$, and this is the central 
point of the method.
First, from Eqs. (20) and (21), the quantity $R(s,b)$
can be evaluated

\begin{equation}
R(s,b)=\ln[{1\over1-\Gamma_{R}(s,b)}]-\Gamma_{R}(s,b),
\label{D}
\end{equation}
and also the errors, $\Delta R(s,b)$, through error propagation from 
$\Delta \Gamma_{R} (s,b)$. We then generate a set of numerical points with
the corresponding errors and making use of the CERN-MINUIT routine this 
set of points with errors, $R(s,b) \pm \Delta R(s,b)$, was fitted by a 
sum of gaussians

\begin{equation}
R_{\textrm{fit}}(s,b)=\sum_{j=1}^{6} A_{j}e^{-B_{j}b^{2}}.
\label{dparamet}
\end{equation}

A typical result is displayed in Fig. 9.
With this par\-ametriza\-tion, $R(s,q)$ in Eq. (23) can be analytically
 evaluated and also the errors, $\Delta R(s,q)$, may be estimated through the
 propagation of the errors from $A_{j}$, $B_{j}$, as given by the routine. 
At last, Eq. (22) leads to $\chi_{I}$(s,q) and the error
 propagation provides $\Delta\chi_{I}(s,q)$. This method was used in Ref. 
\cite{fbv} 
in order to determine the eikonal $\chi_{I}$(s,q). The novel aspect 
of our approach is its use in the estimation of uncertainties.

\begin{figure}
\resizebox{0.48\textwidth}{!}{\includegraphics{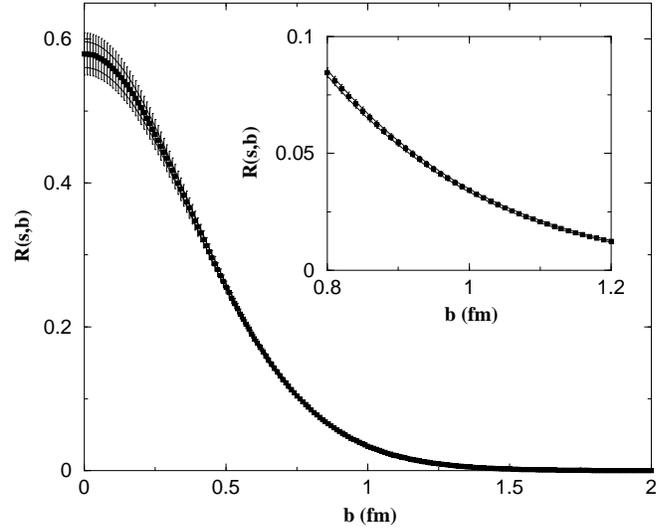}}
\caption{Parametrization for the generated remainder $R(s,b)$ 
by means of Eq. (25), from $pp$ scattering at 62.5 GeV.}
\label{fig:9}       
\end{figure}

\subsection{Zeros and the eikonal at $q^2$ = 0}

In this Subsection, we concentrate on two aspects of the eikonal in the
momentum-transfer space: the existence of zeros (change of sign) and
the results for $\chi_{I}(s,q=0)$ from $pp$ and $\bar{p}p$ scattering.
Here we only present the results and stress some aspects, postponing  
discussions and physical interpretations to the next Section.

In order to investigate the position of the zeros and, mainly, to
determine the uncertainties in its values, we follow Ref. \cite{fbv} and consider
the expected behavior of $\chi_{I}$ at large $q^2$, namely 
 $\chi_{I} \sim q^{-8}$.
In Figs. 10 to 14 we plot the quantity $q^{8}\chi_{I}(s,q)$ as function of $q^2$
for several sets analyzed and obtained with both the numerical and 
the semi-analytical
methods. 

The results from $pp$ scattering with ensembles A and B are shown in Fig. 10 in 
the case of the numerical method and in Fig. 11 with the semi-analytical method.
In the last case, the shaded areas correspond to the uncertainties
obtained from error propagation.

\begin{figure}
\resizebox{0.48\textwidth}{!}{\includegraphics{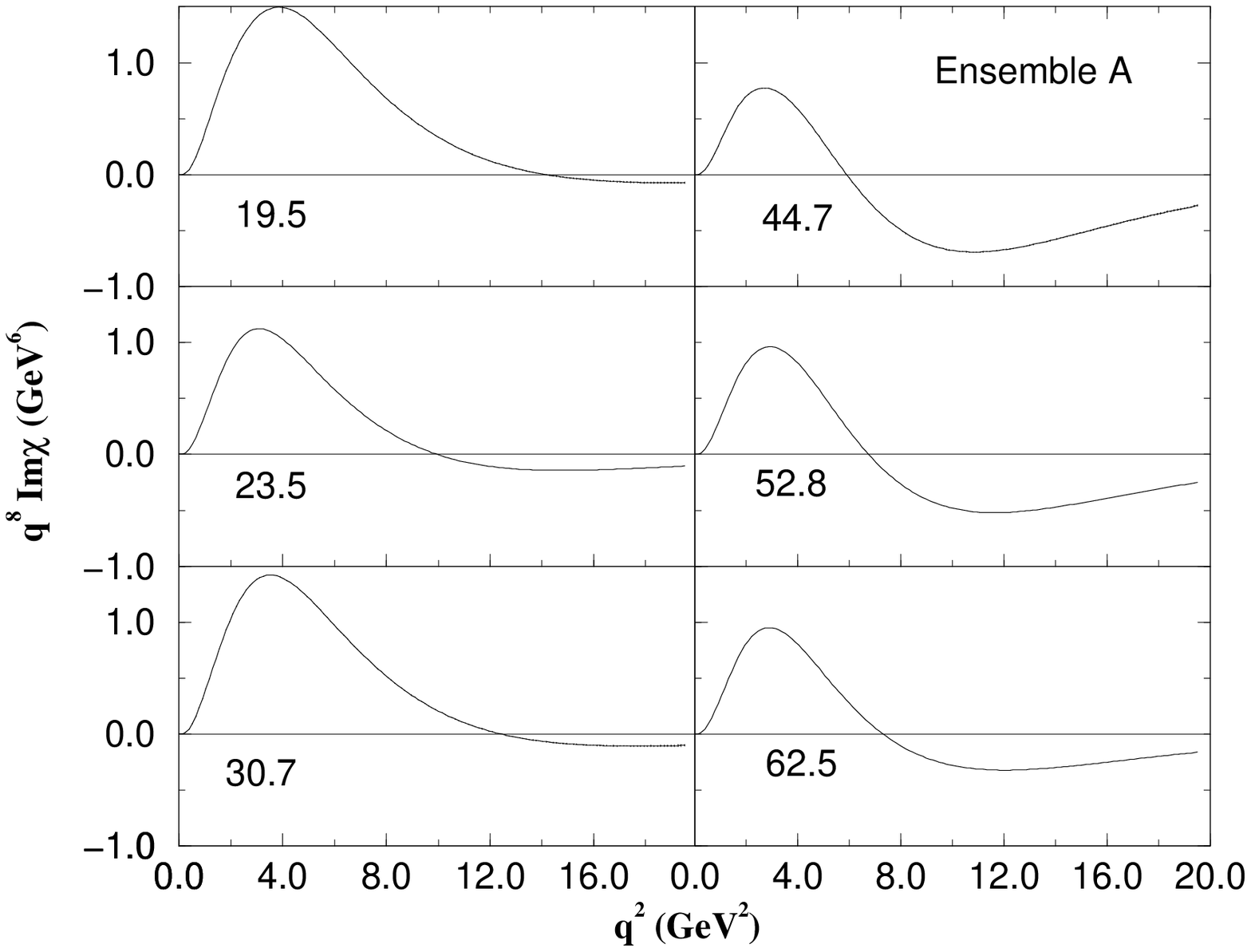}}
\resizebox{0.48\textwidth}{!}{\includegraphics{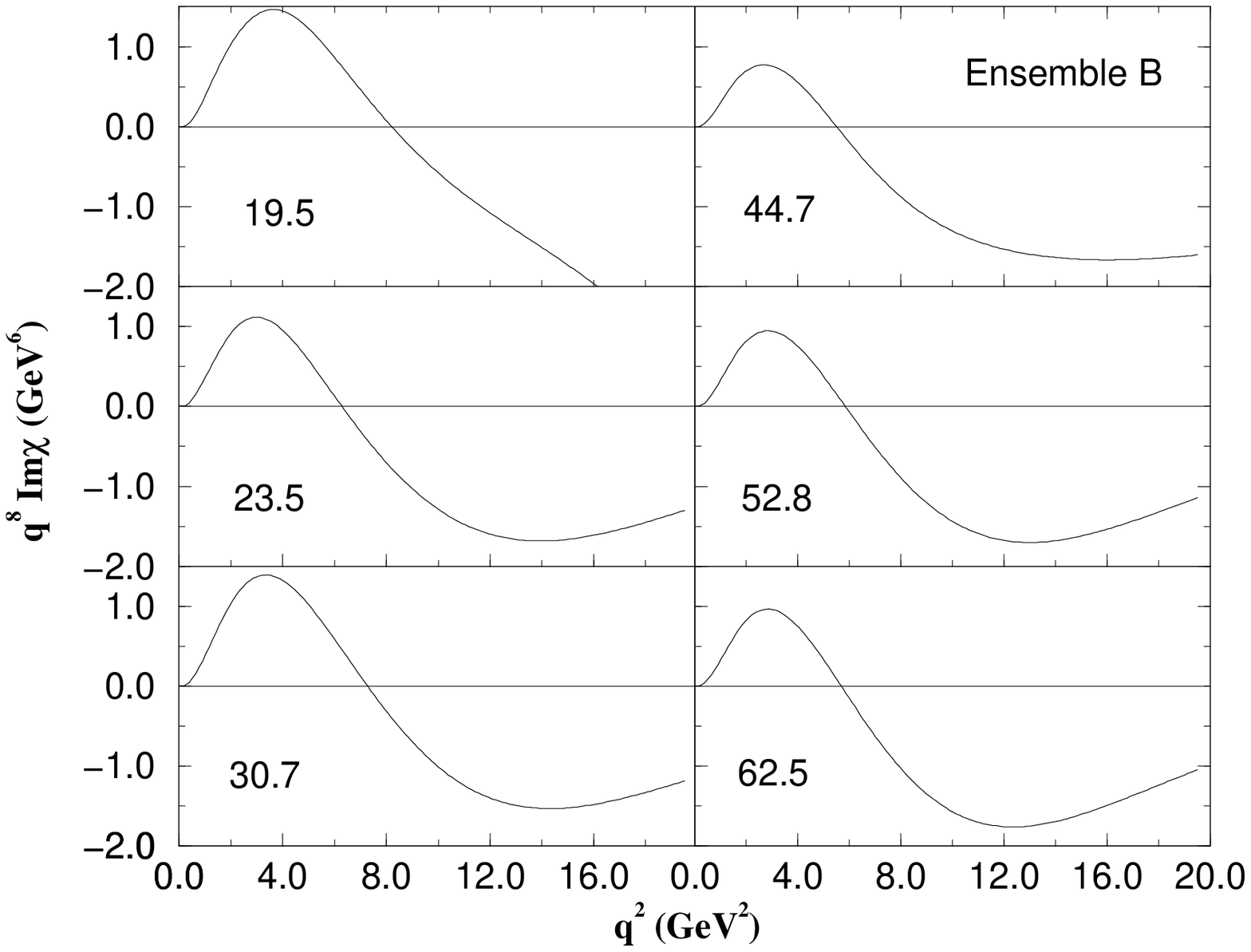}}
\caption{Imaginary part of the Eikonal in the momentum transfer space
 multiplied by $q^8$,
obtained by means of the 
numerical method and ensembles A and B for $pp$ scattering.
The numbers refer to the center-of-mass energy in GeV.}
\label{fig:10}       
\end{figure}

\begin{figure}
\resizebox{0.48\textwidth}{!}{\includegraphics{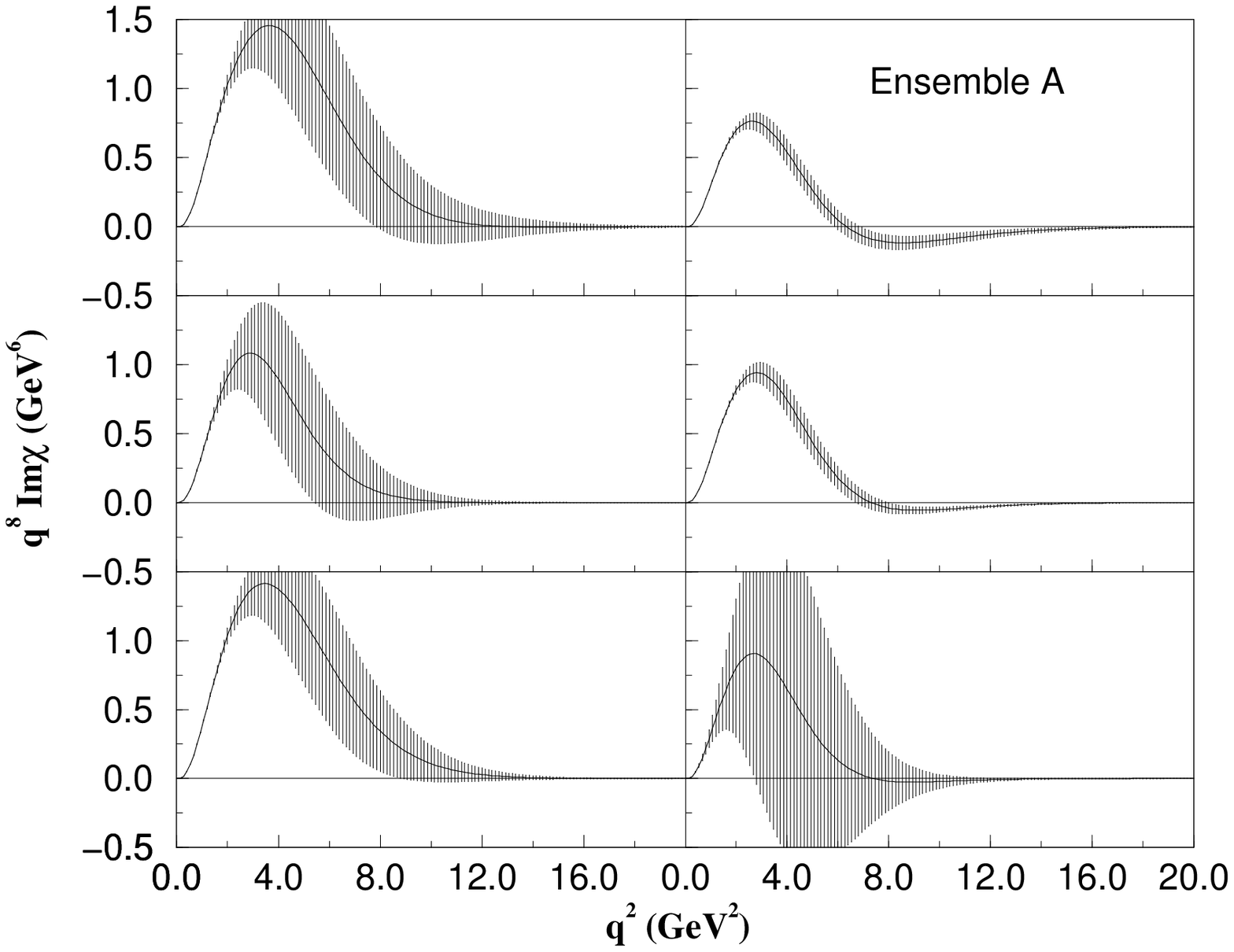}}
\resizebox{0.48\textwidth}{!}{\includegraphics{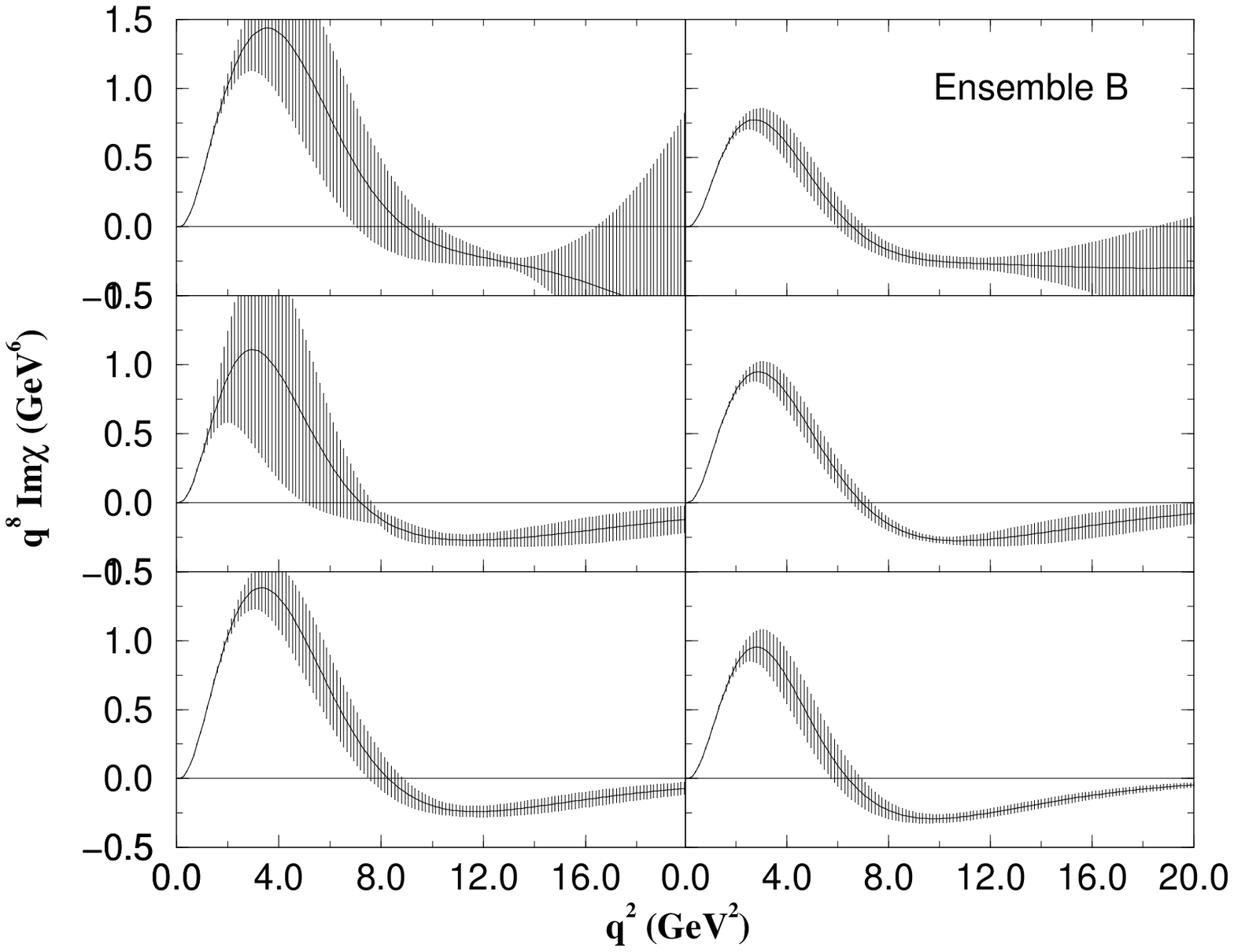}}
\caption{Imaginary part of the 
Eikonal in the momentum transfer space multiplied by $q^8$,
obtained by means of the 
semi-analytical method and ensembles A and B for $pp$ scattering.
The energies are the same as indicated in Fig. 10.}
\label{fig:11}       
\end{figure}

Figure 11 shows clearly the role and the effect of data at large values of
the momentum transfer. In fact, within the uncertainties, ensemble A
shows evidence for the change of sign only at $\sqrt s$ = 44.7 GeV
and 52.8 GeV, which correspond to the sets with the largest
$q^2$ interval with available data (see Fig. 3). On the other hand,
with ensemble B, we find statistical evidence for the change of sign
at all the energies investigated. From these plots we can determine
the position of the zeros and the associated errors from the extrems
of the uncertainty region (in general not symmetrical). The position
of the zero can also be obtained from the numerical method (Fig. 10),
but without uncertainties.

In the case of $\bar{p}p$ scattering the semi-analytical method does not
provide any evidence for change of sign as illustrated in Figs 12, 13 and 14.
As mentioned before, the $pp$ data at $\sqrt s$ = 27.5 GeV are not statistically
consistent with the $\bar{p}p$ data and, for this reaction, the regions with
available data are very small, $q^2 <$ 4.5 GeV$^2$.

\begin{figure}
\resizebox{0.48\textwidth}{!}{\includegraphics{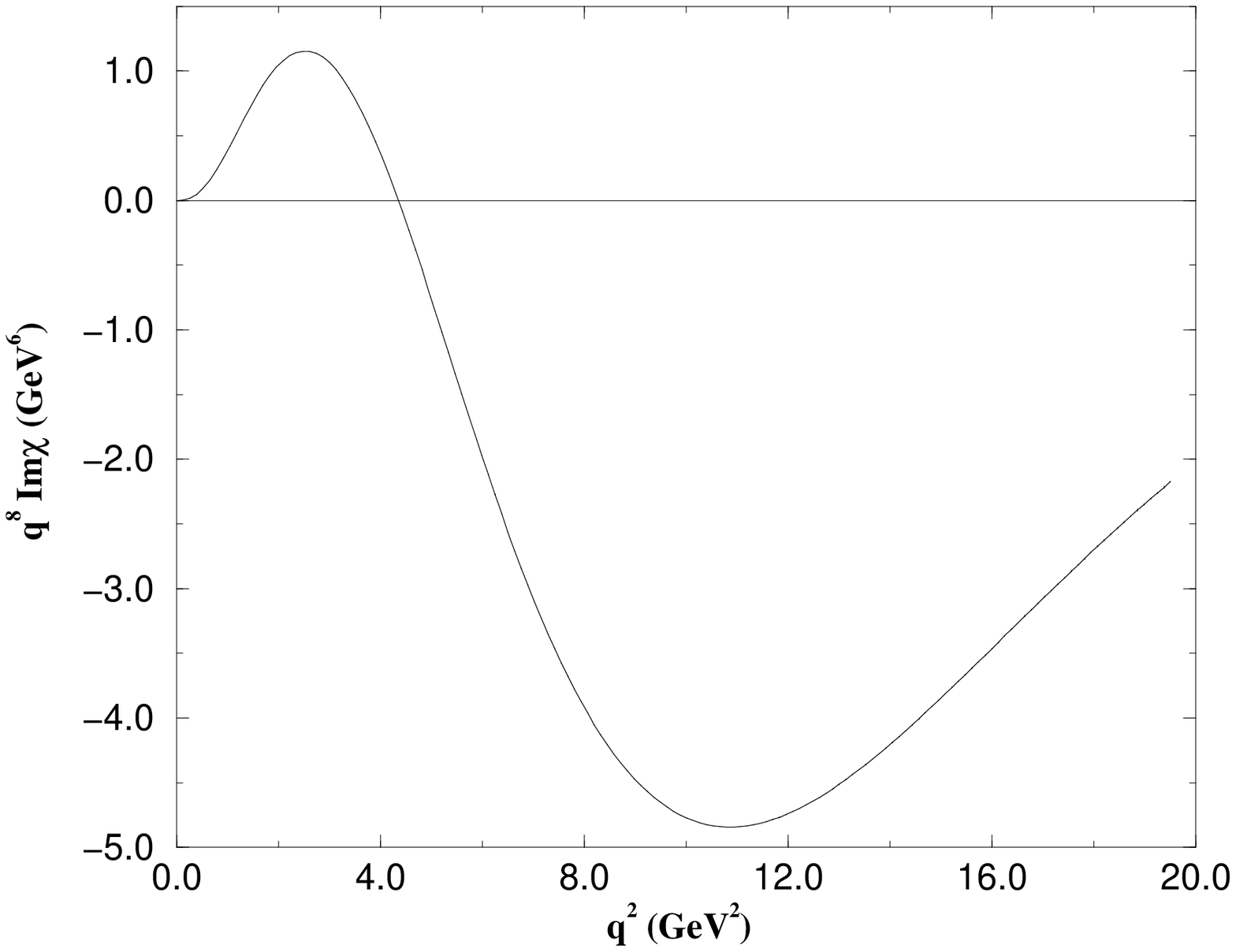}}
\resizebox{0.48\textwidth}{!}{\includegraphics{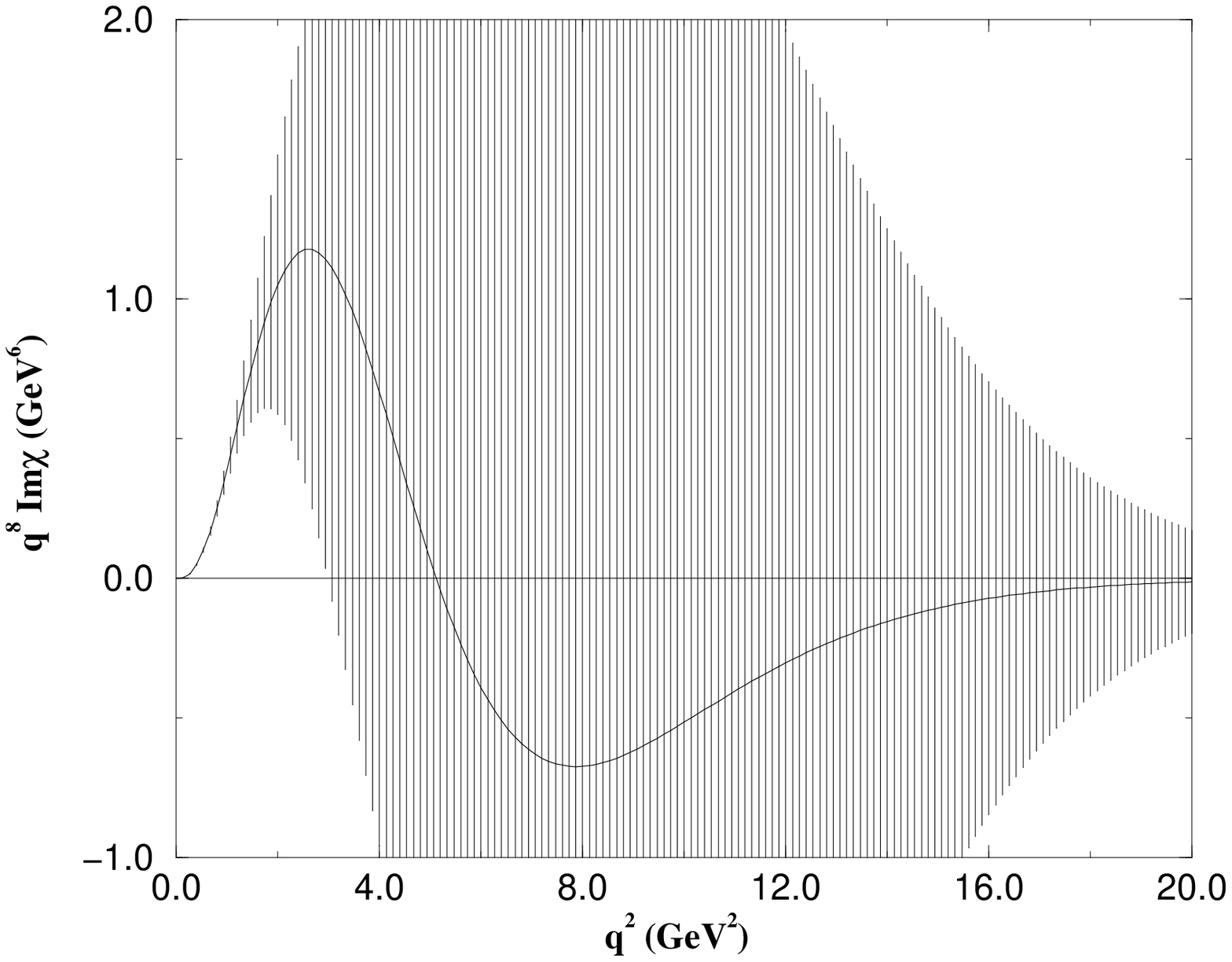}}
\caption{Imaginary part of the
Eikonal in the momentum transfer space multiplied by $q^8$,
obtained by means of the numerical and the
semi-analytical methods with ensemble A for $\bar{p}p$ scattering
at 19.5 GeV.}
\label{fig:12}       
\end{figure}

\begin{figure}
\resizebox{0.48\textwidth}{!}{\includegraphics{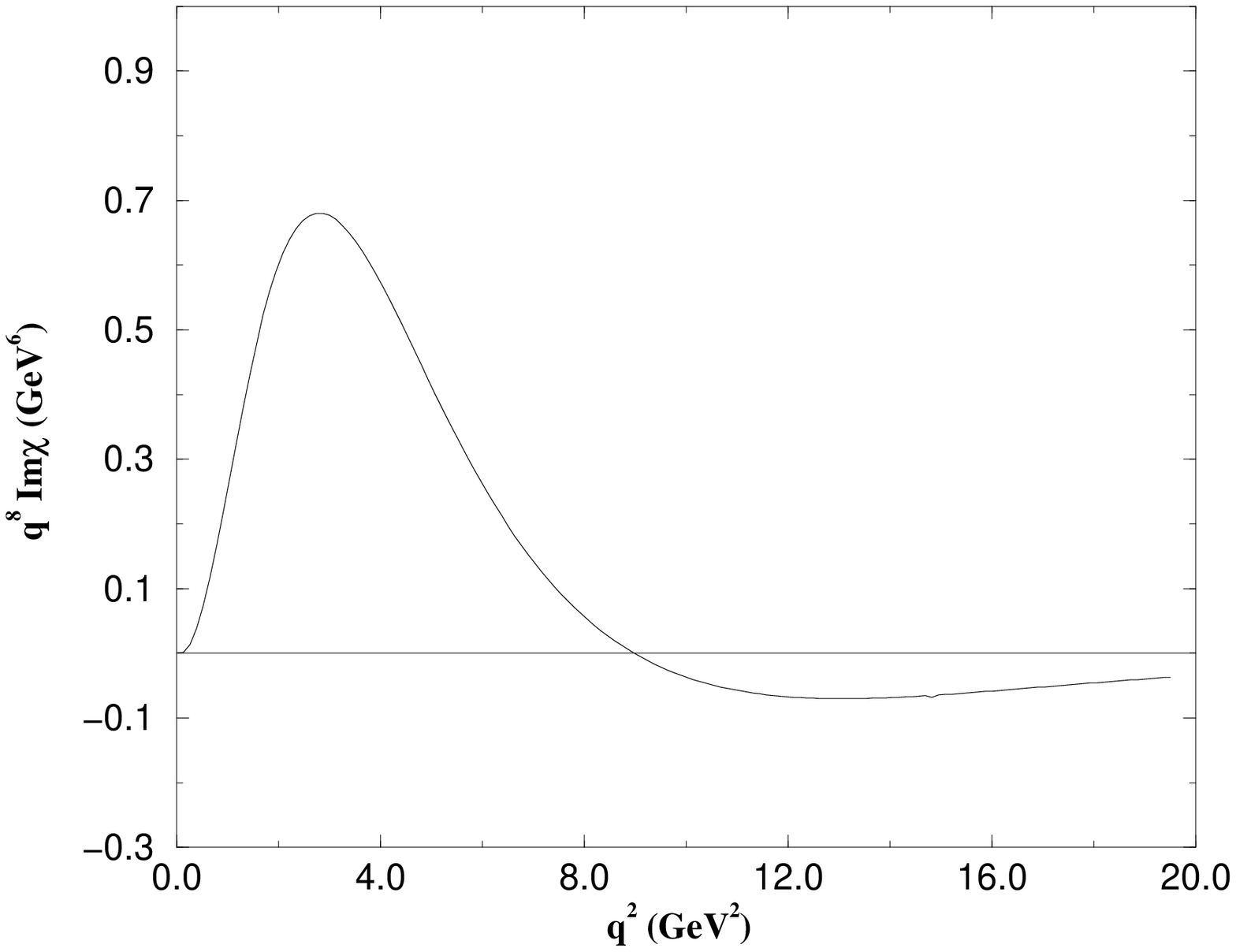}}
\resizebox{0.48\textwidth}{!}{\includegraphics{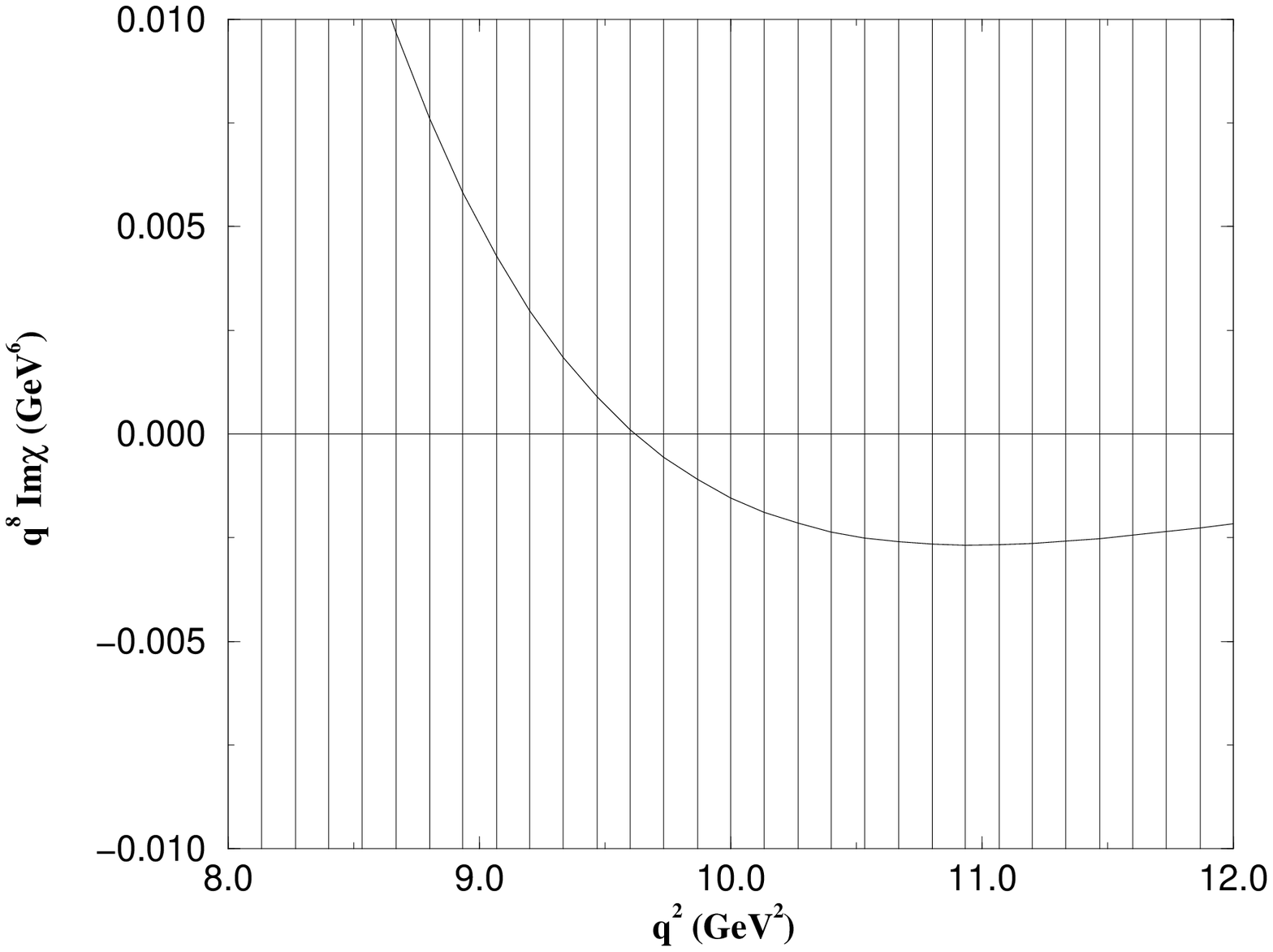}}
\caption{Imaginary part of the
Eikonal in the momentum transfer space multiplied by $q^8$,
obtained by means of the numerical and the
semi-analytical methods with ensemble A for $\bar{p}p$ scattering
at 53 GeV.}
\label{fig:13}       
\end{figure}

\begin{figure}
\resizebox{0.48\textwidth}{!}{\includegraphics{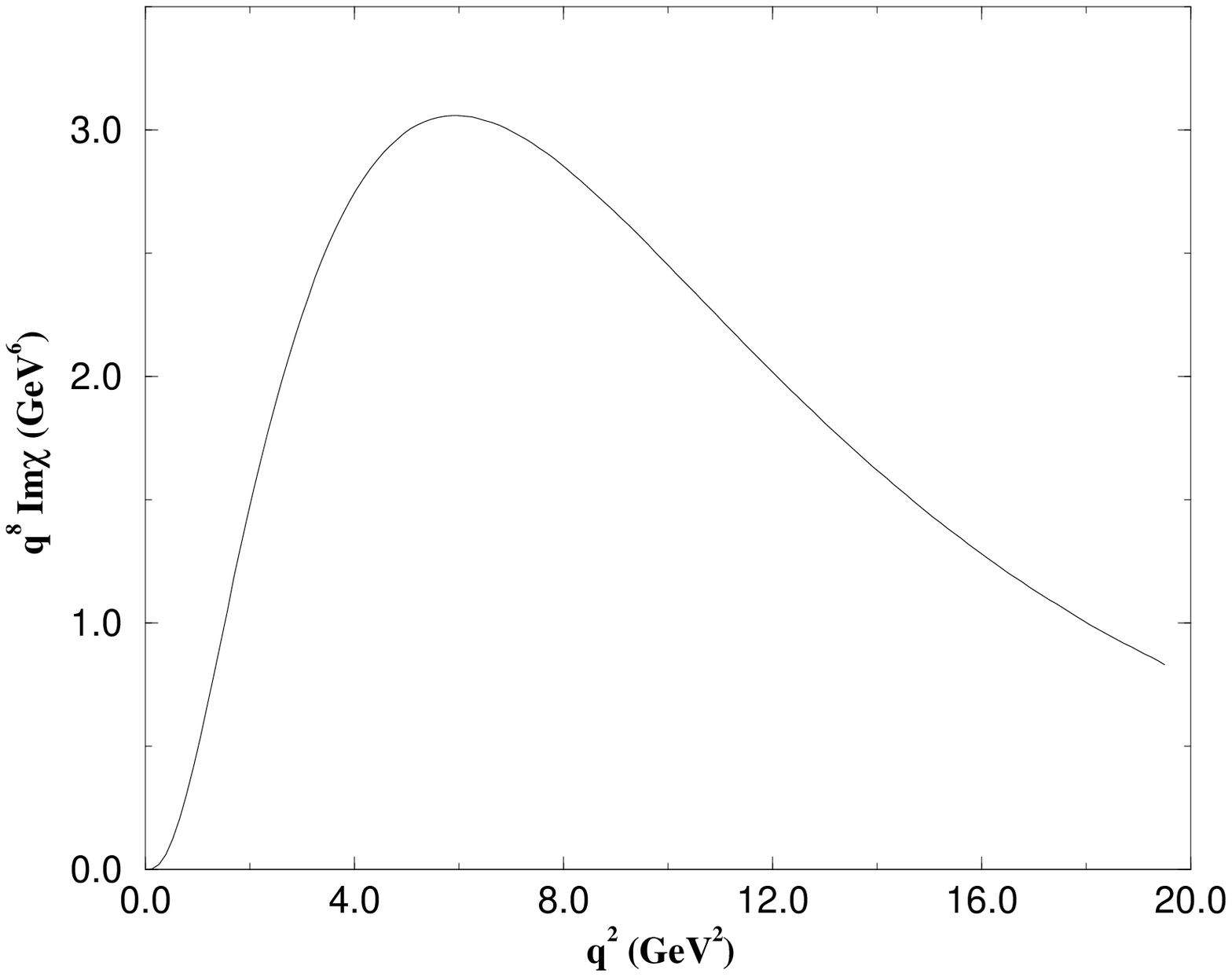}}
\resizebox{0.48\textwidth}{!}{\includegraphics{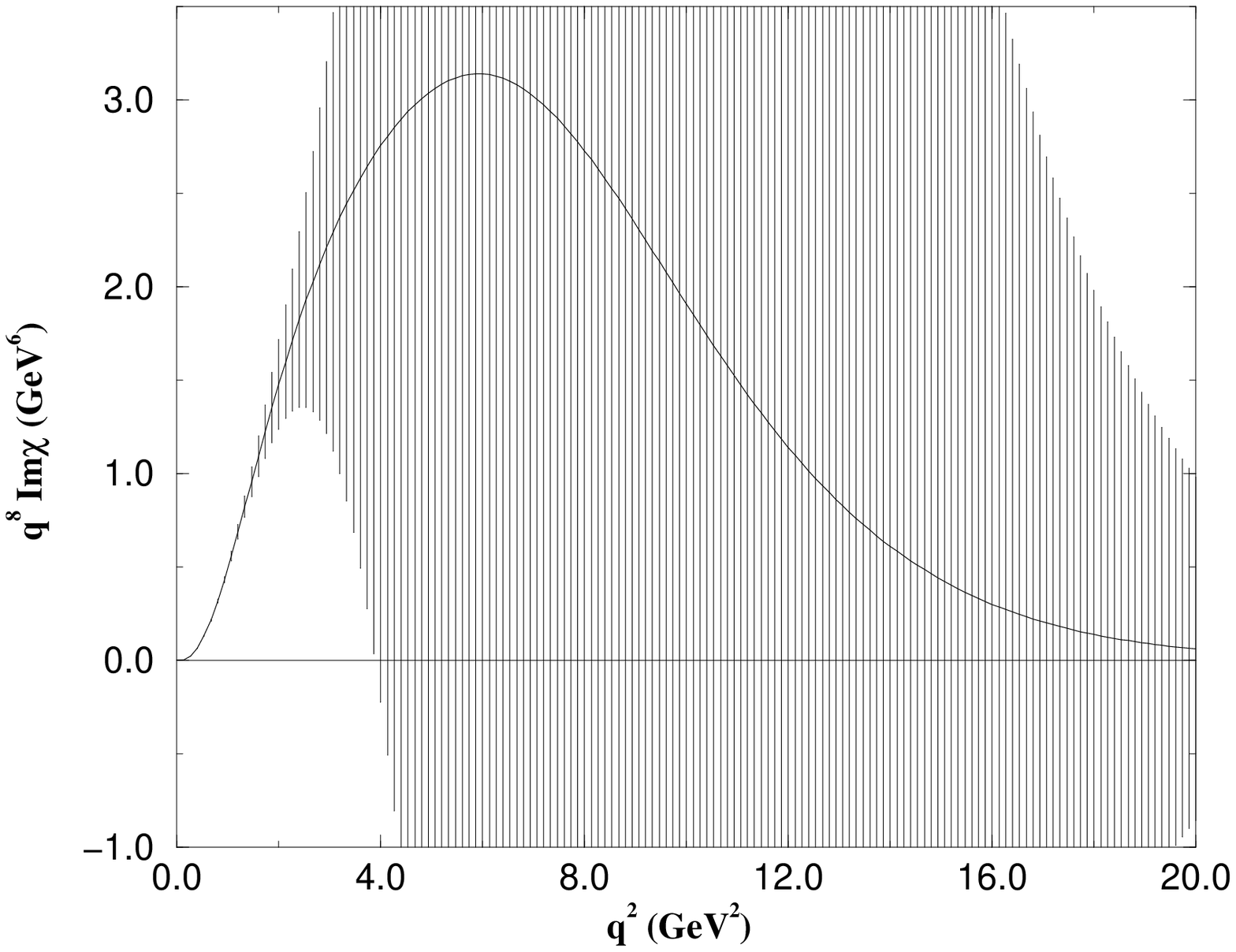}}
\caption{Imaginary part of the
Eikonal in the momentum transfer space multiplied by $q^8$,
obtained by means of the numerical and the
semi-analytical methods with ensemble A for $\bar{p}p$ scattering
at 546 GeV.}
\label{fig:14}       
\end{figure}

We conclude that only the $pp$ data from ensemble B provide statistical evidence
for change of sign. The results obtained with both methods are shown in Fig. 15.
We observe a systematic difference in the position of the zeros as obtained
with the numerical and the semi-analytical methods, the former giving values 
about 0.9 GeV$^2$ below those obtained with the later. As the energy 
increases from
19.5 GeV to 62.5 GeV, the position of the zero decreases from 8.2 to 5.7
GeV$^2$ with the numerical method and from 8.9 to 6.4 GeV$^2$ with the
semi-analytical method. Although the decreasing is not smooth,
the general trend, with both methods, favor that behavior, 
indicating that the position of the zero decreases as the energy increases.

\begin{figure}
\resizebox{0.48\textwidth}{!}{\includegraphics{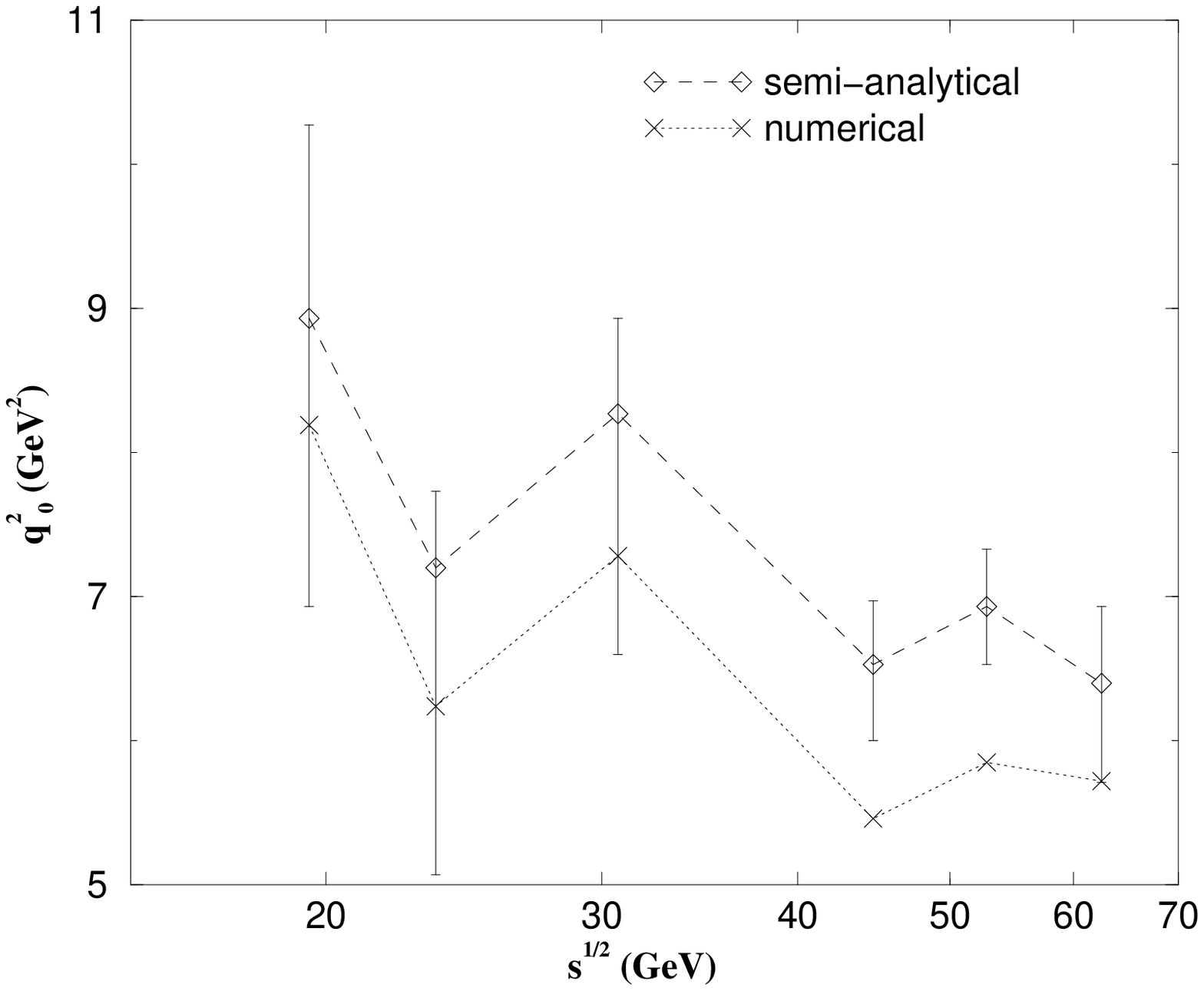}}
\caption{Position of the eikonal zeros in the momentum transfer space
as function of the energy,  
for $pp$ elastic scattering in the interval
$19.5 - 62.5$ GeV: semi-analytical method
(with errors) and numerical method (without errors).
The lines connecting the points have been drawn only to stress the
general trend.}
\label{fig:15}       
\end{figure}

As will be discussed, another quantity of interest is the value of the imaginary
part of the eikonal at $q^2$ = 0. In this case, all the results obtained
with ensembles A and B and through both numerical and semi-analytical methods
are exactly the same for the central values (the numerical method does
not provide the uncertainties). The results from $pp$ and $\bar{p}p$ scattering
are shown in Fig. 16 and will be discussed in Sec. 6.3.
Here we only note that the general dependence of $\chi_{I}(s,q)$ on the
energy from $pp$ and $\bar{p}p$ scattering, is similar to the behavior
of the total cross sections, $\sigma_{tot}^{pp}(s)$ and 
$\sigma_{tot}^{\bar{p}p}(s)$. In fact that is expected since, in first
order, Eq. (1) reads $F_{I}(s,q) \approx \chi_{I}(s,q)$ and, from the
optical theorem, $\sigma_{tot}(s) = 4\pi F_{I}(s, q=0)$. We shall return to
this point in Sec. 6.1.

\begin{figure}
\resizebox{0.48\textwidth}{!}{\includegraphics{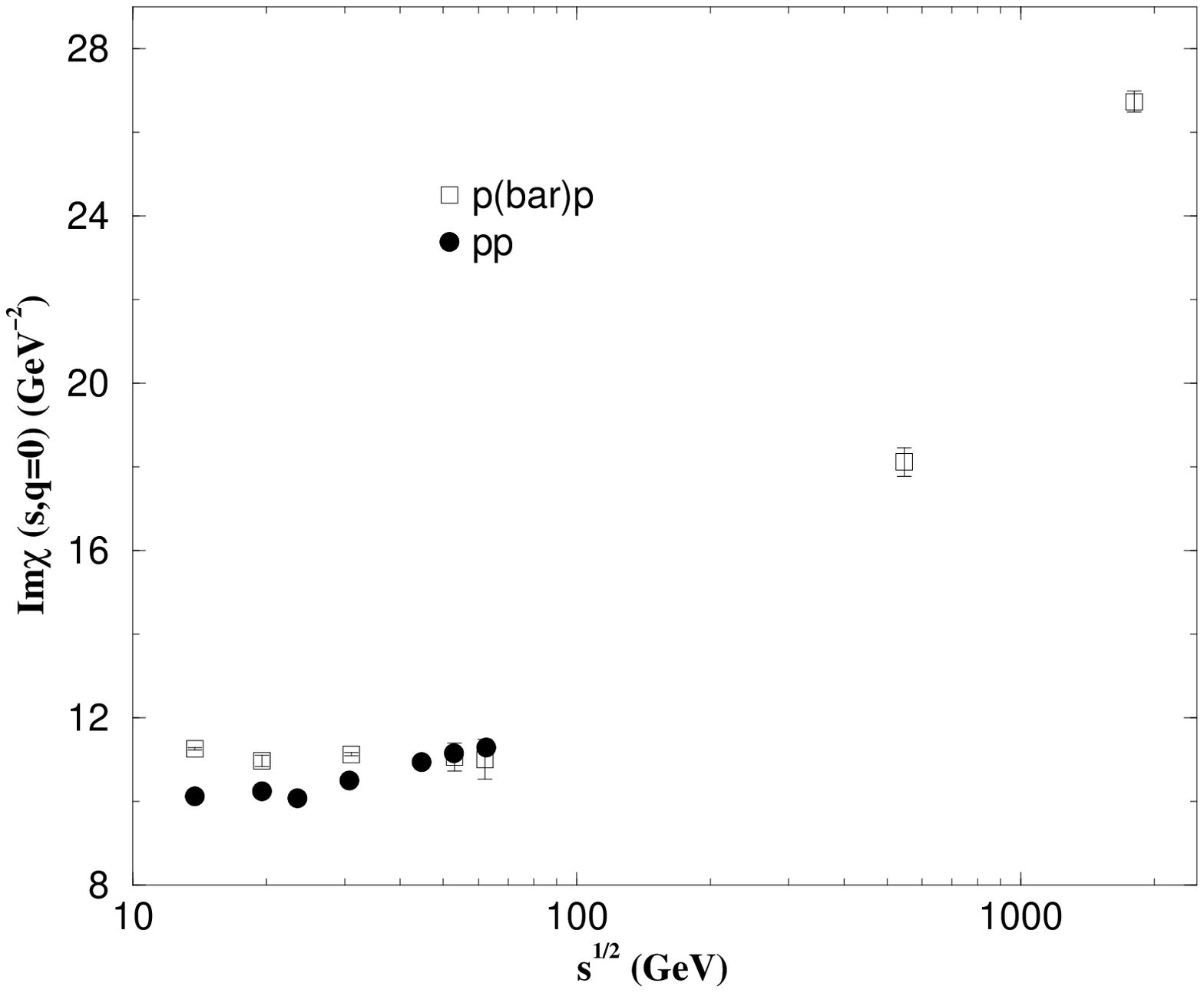}}
\caption{Imaginary part of the eikonal at $q =0$ for
$pp$ and $\overline{p}p$ scattering as function of the energy.}
\label{fig:16}       
\end{figure}

\section{Discussion}
\label{sec:6}

Our main novel results are displayed in Figs. 11, 15 and 16, and
refer, respectively, to the behavior of the eikonal in the $q^2$ - space
(semi-analytical method), the position of the zero as function of
the energy and the eikonal at $q^2$ = 0. In this Section we discuss the
applicability of these results in the phenomenological context,
in connection with those obtained by other authors.

First, we must note that to find a clear dynamical origin for these
``empirical" properties of the eikonal is obviously a very difficult
task. For that reason, we shall consider here a particular
framework, which, despite its simplicity, is suitable for the kind
of points we are interested to raise. We shall refer to the Multiple
Diffraction Theory by Glauber \cite{glauber,gv}, the Chou-Yang model
\cite{chouyang} and the impact picture by Bourrely, Soffer and
Wu \cite{bsw,bsw80,bsw02}. We understand that the essential ideas
to be discussed may be extended to more realistic or general approaches
and, in fact, as we shall show, some limited connections with nonperturbative
QCD may also be inferred.
The discussion will be based on the above 
three aspects, namely the eikonal zeros, the eikonal
at large $q^2$ and the eikonal at $q^2$ = 0.

\subsection{Eikonal zeros}

Let us discuss our results concerning the evidence of the eikonal
zero and its dependence on the energy in the phenomenological context.
We stress that
we shall consider simple approaches, so that some important questions
may be raised and/or discussed without the influence of ``technical"
details.

Let us consider the Multiple Diffraction Theory by Glauber. For the 
scattering between hadrons $A$ and $B$ the eikonal is given
by \cite{glauber,gv,m03}

\begin{eqnarray}
\chi(s,q) = \sum_{i=1}^{N_A} \sum_{j=1}^{N_B} G_A G_B f_{ij},
\nonumber
\end{eqnarray}
where $G_A$ and $G_B$ are the \textit{hadronic} form factors, 
$N_A$ and $N_B$ the number of constituents in each hadron and 
$f_{ij}$ the individual \textit{elementary} scattering amplitudes between
constituents (parton-parton scattering amplitudes).
 If we consider, for simplicity, that the elementary amplitudes
are all the same, denoted by $f$, and that $N_A N_B \equiv N$, for $pp$ 
scattering we have for the imaginary parts

\begin{eqnarray}
\chi_I (s,q) = N G_{p}^2\ f_I.
\end{eqnarray}

This expression indicates that, in principle, the zero in the 
imaginary part of the eikonal may
be associated either with the form factor or with the elementary amplitude. 
Let us discuss both possibilities.

In the phenomenological context (multiple diffraction models), the
interpretation of the zero as associated with the elementary amplitude has
been discussed, for example, in \cite{m03}. On the other hand, more recently,
elementary amplitudes have been determined from nonperturbative QCD
\cite{mmt,mm03},
by means of the Stochastic Vacuum Model \cite{svm}. The fundamental input
is the gluon gauge-invariant two-point correlation functions, which
eventually determines the structure of the elementary amplitude.
Two variants may be found in the literature, which can be distinguished
by the behavior of the correlators in the region of small (physical)
distances. From lattice QCD, in both quenched approximation (absence of
fermions) and full QCD (dynamical fermions included), the parametrized
correlator has a divergent term, $1/x^4$, at the origin \cite{digi}. In
contrast, the parametrization introduced by Dosch, Ferreira and Kr\"amer
is finite at the origin \cite{dosch}. It has been shown in \cite{mm03}
that the elementary amplitudes determined with the lattice parametrization
are characterized by a monotonic decrease of the amplitude with the
momentum transfer, through positive values (no zeros). On the other hand,
the finite correlator, by Dosch, Ferreira and Kr\"amer, leads to an elementary 
amplitude which presents a
zero at $q^2 \approx$ 0.5 GeV$^2$ (the position of the zero depends on
the value of the gluonic correlation length) and then goes asymptotically to
zero through negative values \cite{mmt}. It should be noted that this
formalism is intended for asymptotic energies ($s \rightarrow \infty$) and
small momentum transfer (typically $q^2 \leq {\cal O}(1)$ GeV$^2$),
which put some limitations in the conclusions that may be inferred,
as discussed in detail in \cite{mmt,mm03}.
It has been claimed that the divergent term in the correlator is a
perturbative effect that should not be included in nonperturbative
calculations. However, since there is no conclusive answer to this 
question in the literature, the possibility that the elementary amplitudes
have no zeros can not be disregarded.

Let us, therefore, consider the possibility that the eiko\-nal zero is 
associated
with the form factor, that is, assuming the positivity of the elementary
amplitude as in the case of the parametrization from
lattice QCD \cite{mm03}. 
In that case, a striking result
is the dependence of the position of the eikonal zero on the energy. From a
``pragmatic" or ``empirical" point of view, for $pp$ and $\bar{p}p$ hadronic
elastic scattering, that dependence may be associated with the shrinkage of the
diffraction peak, an effect experimentally verified when the energy increases
 in the
region 23 GeV $\leq \sqrt s \leq$ 1.8 TeV \cite{matthiae}. In fact,
as recalled before, in first order, $F_{I}(s,q) \approx \chi_{I}(s,q)$ and,
as it is known, the diffraction peak is dominated by the imaginary part
of the amplitude.
This possibility brings novel insights, since the main point is the implication
in hadronic form factors depending on the energy, which is an old phenomenological
conjecture. Despite of limited theoretical foundation, it has been shown, in the
past, that the introduction of energy dependence in form factors leads
to good descriptions of the experimental data \cite{gofs,bcmp}. In particular,
$pp$ and $\bar{p}p$ elastic scattering data can be well described by means
of both geometrical models \cite{menon} and hybrid Regge-dual models \cite{covo}.
We understand that to explore this dependence in well founded theoretical
grounds (with analyticity and crossing taken explicitly into account)
may lead to novel results in the investigation of the soft diffractive processes
in general.

As commented in \cite{kawa2}, the eikonal zero may indicate the
existence of two components in the hadronic elastic process, one
dominating the region of small momentum transfer (long range)
and the other the region of large momentum transfer (short range).
Roughly, these components could be associated with the regions before
and after the position of the zero, respectively. If that is the case
the characteristic of the long range component is the positivity of
the eikonal in the $q^2$ - space, in contrast with its negative values
in the short range region. We shall discuss this later case in the
following Subsection.

All the previous discussion was based on high-energy hadronic interactions
($pp$ elastic scattering) and, therefore, the form factor concerns the hadronic 
structure of the proton. To end this subsection, let us discuss some results 
recently obtained from experiments on elastic electron-proton scattering
and related to the electromagnetic structure of the proton. Although, 
presently, discrepancies from two different experimental
techniques characterize the results, future experimental and
theoretical developments may bring new insights on possible connections
between hadronic and electromagnetic form factors, as discussed in
what follows.

The experiments on elastic $e-p$ scattering provide information on the ratio
$R \equiv \mu_p G_{Ep}(q^2)/G_{Mp}(q^2)$, where $G_{Ep}(q^2)$ and
$G_{Mp}(q^2)$ are the Sachs electric and magnetic form factors and
$\mu_p$ the proton magnetic moment. The traditional method used to extract 
the form factors is based on the Rosenbluth separation technique
\cite{rose} and fits to data on $R$, as function of the momentum
transfer, have yielded a scaling behavior: $R \approx 1$ \cite{scaling1}.

Recently, this ratio has been measured at the Jefferson Laboratory
by means of the polarization transfer technique
($ \vec{e} p \rightarrow e \vec{p} $) and the results indicated a decrease
of the ratio from 0.97 to 0.27, as the momentum transfer increases from
0.5 to 5.5 GeV$^2$ (nonscaling behavior) \cite{jones,gayou1,gayou2}.
Extrapolation from empirical fits indicates a zero (change of
sign) at $q^2 \approx $ 7.7 GeV$^2$ \cite{gayou2} and a recent phenomenological
description of these data, in the context of a Regge parametrization for
Generalized Parton Distributions yielded a zero in the electric
form factor at  $q^2 \approx $ 8 GeV$^2$ \cite{guidal}. Certainly, these
results suggest some possible connections between the inferred
position of the zero in the electric form factor and the position
the eikonal zeros at $q^2 \approx $ 6 - 9 GeV$^2$ (Figure 15), which,
as discussed before, may be associated with the hadronic form factor.

However, new measurements performed at Jefferson Lab through the Rosenbluth 
technique have confirmed the scaling behavior \cite{christy} and the
same result has been obtained in more recent and improved
measurements \cite{qattan}: global analysis of the cross section data
indicates  $R \approx 1$ in the momentum-transfer region $q^2$: 0 - 6
GeV$^2$.

As discussed in \cite{qattan}, despite the inconsistency of the
results in the region of large momentum transfer, both measurements
(polarization transfer and improved Rosenbluth) are of
comparable precision and the origin of the discrepancy is not clear yet
(see \cite{qattan} for references on possible theoretical corrections
in development).

Certainly this enigma must be explained before any attempt to conclude on
the existence or not of a zero in the electric form factor
at $q^2 \approx $ 7 - 8 GeV$^2$. However, in case that further
experimental and/or theoretical developments might favor
the polarization-transfer form factors \cite{qattan}, it may be important
to investigate possible connections between zeros in the electric and 
hadronic form factors.

\subsection{Eikonal at large $q^2$}

Figure 11 (and also 10) shows another interesting aspect of the
extracted eikonal, in the region of intermediate and large momentum transfer.
The results indicate that, after the zero, the eikonal reaches a minimum 
and then
approaches zero through negative values. That may indicate a second asymptotic
 zero,
which is obviously expected as $q^2 \rightarrow \infty$. However,
this approximation to zero through negative values brings some new insights
in model constructions.

For example, let us return to the impact picture by Bourrely, Soffer and Wu.
In this model, the contribution from the Pomeron exchange is assumed to be 
factorized in $s$ and $b$, with the impact parameter dependence given
by the parametrization (11) for the form factors in the $q^2$ - space.
Since, as $q^2 \rightarrow \infty$, $G(q) \rightarrow$ 0 and
$f_{\mathrm{BSW}}(q) \rightarrow -$ 1, this parametrization qualitatively
reproduces the change of sign in the eikonal and the limit to
zero through negative values.
However, as quoted before, the position of the zero fixed at $a^2 =$
3.45 GeV$^2$ \cite{bsw02} is in disagreement with our extracted behavior.

It should be noted that this model 
overestimates the differential cross section data in the region
of large momentum transfer
\cite{bsw,bsw02}, a point also raised in \cite{kawa2}. If this effect
is not only a consequence of the above factorized Pomeron contributions, it may
be associated with the behavior of the eikonal after the zero, as
mentioned before, that is, the short range contribution. In this case,
it may be recalled that    
a \textit{modified} BSW factor (mBSW), given by

\begin{eqnarray}
f_{\mathrm{mBSW}}(q) = \frac{1 - q^2/a^2}{1 + q^4/a^4},
\end{eqnarray}
can also generate the above behavior of the eikonal and leads to good
descriptions of the experimental data at large momentum transfer.
This function has been introduced in \cite{bcmp} and
used in both the geometrical approach
(with $a^2$ = 8.2 GeV$^2$) \cite{menon} and a hybrid Regge-dual
model \cite{covo}. The differences between these two ansatz are illustrated
in Fig. 17 for the position of the zero at $a^2 =$ 3.45 GeV$^2$.

We understand that tests  with both functions, taking into account the
possibility that the parameter $a^2$ depends on the energy may lead
to improved descriptions of the experimental data. We are presently 
investigating this subject.

\begin{figure}
\resizebox{0.48\textwidth}{!}{\includegraphics{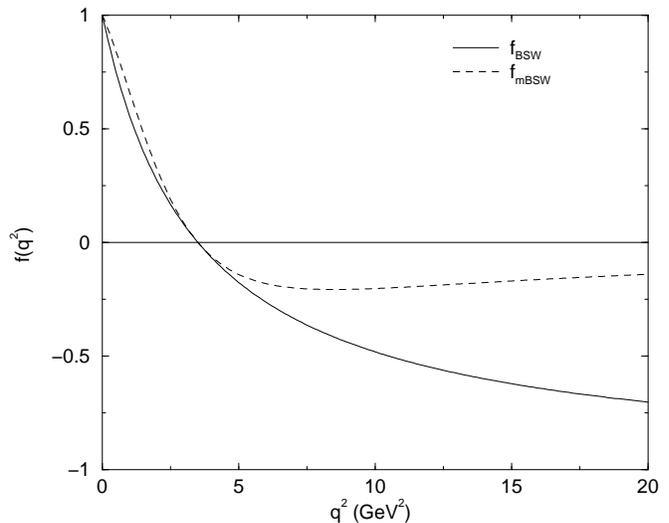}}
\caption{The Bourrely-Soffer-Wu term, from Eq. (11), and the modified form,
Eq. (28), as function of the momentum transfer and for
$a^2$ = 3.45 GeV$^2$.}
\label{fig:17}       
\end{figure}

\subsection{Eikonal at $q^2$ = 0}

Figure 16 shows the imaginary part of the eikonal at $q^2 = 0$, as
obtained with both the numerical and semi-analytical methods.
As in the previous Subsections, we discuss here some simple
examples concerning the applicability
of these results in the phenomenological context.

Let us return to the Glauber model, expressed by Eq. (26).
Since at $q^2 = 0$ the hadronic form factors are normalized by
$G_p$ = 1, the dependence on the energy from the imaginary part
of the eikonal at $q^2 = 0$ may be associated either with the number of
 participants
constituents or with the elementary amplitude, or both. Only to treat a 
toy example,
let us assume $N = N_A$ x $N_B$ fixed at 3\ x\ 3 = 9 and consider the 
Optical Theorem
at the elementary level, namely 
$\sigma_{elem}(s) = 4\pi f_I(s, q=0)$. In that case
we can express the elementary cross section in terms of the
imaginary part of the eikonal:

\begin{eqnarray}
\sigma_{elem}(s) = \frac{4\pi}{N} \chi_I(s, q=0).
\end{eqnarray}

The values of the elementary cross sections, obtained through this formula,
with $N$ = 9,
from the extracted values of the eikonal at $q^2$ = 0, are
displayed in Table 4. We see that even with this toy model
and for fixed $N$ = 9, the elementary cross sections at the
ISR energy region, from analysis of both $pp$ and
$\bar{p}p$ scattering, is of the order of 5 - 6 $mb$,
a reasonable estimation of the partonic cross sections.
Parametrizations of these cross sections as function of
the energy have been discussed in \cite{cmmm}.

In this simple example, we considered 
the number $N$ of participants in the elementary scattering
as fixed.
However, in principle, this number may also depend on the energy
and, for example, for a fixed elementary cross section
(typically 5 - 6 mb) the increase in $\chi_I(s, q=0)$ may
be associated with an increase of $N$. These considerations are aimed only to 
exemplify some possible uses of the
extracted eikonal. To go on with this discussion demands, however, a more
realistic formalism.

\begin{table}
\begin{center}
\caption{Elementary (parton-parton) cross sections from $pp$ and
$\bar{p}p$ scattering, Eq. (28). The errors come from the uncertainties in the
extracted values of $\chi_I(s, q=0)$ through the semi-analytical method. }
\label{tab:4}
\begin{tabular}{lll}
\hline\noalign{\smallskip}
$\protect\sqrt{s}$ & $\protect\sigma_{elem}^{pp}$ & 
$\protect\sigma_{elem}^{\bar{p}p}$ \\
(GeV) & (mb) & (mb) \\
\noalign{\smallskip}\hline\noalign{\smallskip}
13.8 & 5.50$\pm$0.05 & 6.12$\pm$0.02 \\ 
19.4 & 5.57$\pm$0.02 & 5.96$\pm$0.08 \\ 
23.5 & 5.48$\pm$0.07 & - \\ 
30.7 & 5.71$\pm$0.03 & 6.05$\pm$0.02 \\ 
44.7 & 5.95$\pm$0.03 & - \\ 
52.8 & 6.07$\pm$0.02 & 6.01$\pm$0.18 \\ 
62.5 & 6.14$\pm$0.07 & 5.99$\pm$0.26 \\ 
546.0 & - & 9.85$\pm$0.18 \\ 
1800.0 & -  & 14.53$\pm$0.14 \\
\noalign{\smallskip}\hline
\end{tabular}
\end{center}
\end{table}

\section{Conclusions and final remarks}
\label{sec:7}

In this work we have presented the results of analytical fits
to $pp$ and $\bar{p}p$ differential cross section data, in a
model independent way. As explained, we were not interested in
the dependence of the fit parameters with the energy, but
only in the best statistical results, in a model-independent 
context.

By means of both a numerical and a semi-analytical methods,
we have determined the imaginary part of the eikonal
in the momentum-transfer space. Based on the confidence
region of the statistical results, we conclude that the
eikonal presents a zero and that the position of
the zero, roughly, decreases from 8.5 GeV$^2$ to
6.0 GeV$^2$ as $\sqrt s$ increases from 20 to 60 GeV. 
After the zero, the eikonal has 
a minimum and then goes to zero through negative values.

We have presented a critical review on several aspects related to
analytical fits to the differential cross section data and
also discussions on the applicability of our results in the
phenomenological context. Although limited to very simple
models, we understand that these aspects may be extended
to more realistic approaches.

As discussed, in the phenomenological context, the positivity of 
the elementary amplitudes determined from quenched and full QCD 
\cite{mm03} suggests
that the eikonal zero might be associated with the hadronic
form factor. In that case, the decrease in the position of the
zero as the energy increases (Figure  15) implies in hadronic
form factors depending on the energy. We understand that investigation 
of this possibility on well founded theoretical bases may lead
to important developments in the treatment of the soft 
diffractive processes. 

We have also called the attention to the
fact that the position of the eikonal zeros are consistent with
the results for the electromagnetic form factor obtained
from polarization-transfer experiments. It is expected that
the discrepancies between the Rosembluth and polarization-transfer
form factors may be resolved in the near future \cite{qattan}.

As we have shown, only data at large values of the momentum transfer
can provide precise answers on the topical question related with
the eikonal zeros and their dependence on the energy. 
In that sense, it would be very important if the experiments at RHIC and
LHC could extend the region of momentum transfer to be investigated;
as stated in \cite{kawa2}: ``Such experiments will give much more
valuable information for the diffraction interaction rather than
to go to higher energies".

\begin{acknowledgement}
M.J.M. and A.F.M. are thankful to FAPESP for financial support 
(Contracts No. 01/08376-2, No. 00/04422-7) and P.A.S.C. to
Unipam - Centro
Universit\'ario de Patos de Minas, MG.
We are grateful to R.F. \'Avila, S.D. Campos, E.G.S. Luna, J. Montanha,
and R.C. Rigitano for useful discussions.
\end{acknowledgement}

\end{document}